\documentclass[prd,superscriptsize,reprint]{revtex4-1}
\usepackage{graphicx,xcolor,overpic,mathtools}
\usepackage{amsthm,amsmath,amssymb}
\usepackage{hyperref}
\PassOptionsToPackage{normalem}{ulem}
\usepackage{ulem} 
\usepackage{cleveref}
\usepackage{sidenotes}
\usepackage{bm}
\usepackage{url}
\usepackage{physics}
\usepackage[makeroom]{cancel}
\usepackage{tensor}
\usepackage{mathrsfs}
\usepackage{slashed}
\usepackage[mode=buildnew]{standalone}
\newcommand{\lag}{\mathscr{L}}

\NewDocumentCommand{\evat}{sO{\bigg}mm}{%
  \IfBooleanTF{#1}
   {\mleft. #3 \mright|_{#4}}
   {#3#2|_{#4}}%
}
\usepackage{enumerate}

\newcommand{\eps}{\epsilon}
\newcommand{\rb}{\Bar{r}}
\begin{document}

\title[Quasi-universal relations for generalized Skyrme stars]{Quasi-universal relations for generalized Skyrme stars}

\author{Christoph Adam}
\author{Alberto Garc\'ia Mart\'in-Caro}
\author{Miguel Huidobro}%
\author{Ricardo V\'azquez}
\affiliation{%
Departamento de F\'isica de Part\'iculas, Universidad de Santiago de Compostela and Instituto
Galego de F\'isica de Altas Enerxias (IGFAE) E-15782 Santiago de Compostela, Spain
}%
\author{Andrzej Wereszczynski}
\affiliation{
Institute of Physics, Jagiellonian University, Lojasiewicza 11, Krak\'ow, Poland
}%
\affiliation{IGFAE}

\date[ Date: ]{\today}
\begin{abstract}
First proposed in 2013 by Yagi and Yunes, the quasi-universal \emph{I-Love-Q relations} consist of a set of relations between the moment of inertia, the spin-induced quadrupole moment and the electric quadrupolar tidal deformability of neutron stars which are independent of the Equation of State (EoS) within an accuracy of $\sim1\%$. In this work, we show that these relations hold for different Skyrme-based nuclear matter EoS and also for the star-like solutions of different Einstein-BPS-Skyrme-models, some of which do not even present a barotropic equation of state. Further, other quasi-universal relations are analyzed, and together with recent GW observations, we use them to select the generalized Skyrme model that better reproduces observations. Our results reaffirm both the universality of the \emph{I-Love-Q} relations and the suitability of generalized Skyrme models to describe nuclear matter inside neutron stars.
\end{abstract}
\maketitle
%\begin{minipage}{\textwidth}
\tableofcontents
%\end{minipage}
\section{Introduction}
The Skyrme model \cite{Skyrme61} and its generalizations \cite{JACKSON1985101,Vecmes2,MeissnerZahed,Marleau:1990nh,Adam_2010,Sutcliffe:2010et} consist in a set of relativistic, effective-field-theoretic models of interacting Goldstone bosons which have been proposed to describe strongly interacting matter in a low energy regime. Indeed,  baryons, nucleons and nuclei, whose existence can not be inferred by perturbative QCD methods, are described within the Skyrme models as (topological) solitonic configurations of the underlying bosonic degrees of freedom. During the last few decades, many properties of nucleons \cite{Adkins:1983ya} and nuclei \cite{Braaten:1985np,Weigel:1986zc,Battye:2009ad,Lau:2014baa,Halcrow:2016spb,Naya_2018,Gudnason:2018ysx,Halcrow:2020gbm} have been reproduced using these models. These results have contributed to establish the Skyrme model approach as a well-motivated proposal for the description of nuclear matter. Furthermore, in recent years there has been a growing interest in obtaining self-gravitating solutions of the Einstein-Skyrme system in order to determine whether the Skyrme model and its generalizations are also able to describe the properties of matter inside neutron stars (NS) \cite{NelmesPiette,Canfora_2013,Adam_2015a,Adam_2015b,Naya_2019,Adam:2020yfv}. 

On the other hand, one of the most outstanding challenges of current astrophysical research is to obtain information about the
equation of state of ultra-dense matter from neutron star observations. In particular, apart from their masses and radii, other interesting observable properties of NS are their quadrupole moments, spin angular velocity (angular momentum), and deformability against tidal forces---which is encoded in the so-called Love numbers \cite{Hinderer:2007mb,Postnikov:2010yn}. All these properties can be constrained by their imprints into the waveform of a gravitational wave signal emitted by an inspiraling
binary neutron star system. Indeed, binary NS systems are one of the most promising sources of gravitational waves (GW) within the detection range for second generation observatories, such as Advanced LIGO, Advanced VIRGO \cite{Harry:2018hke,EOSMeasure}, or KAGRA \cite{KAGRA2018}. The observation of GW emitted during the coalescence of the stars in such systems---especially in the last part of the merging, in which the stars are subject to large tidal deformations due to the extremely strong gravitational fields involved---will shed light onto the equation of state (EoS) of matter at very high densities, well beyond the nuclear saturation point. 

A particularly interesting property of compact stars is the apparently universal relation between the moment of inertia, the Love numbers and the quadrupole moment (I-Love-Q relations) of such stars. These I-Love-Q relations, firstly proposed by K. Yagi and N. Yunes in \cite{Yagi_2013}, when applied to NS, allow to break the degeneracy between the quadrupole moment and the NS spins in the gravitational waveforms of inspiraling NS binaries. Therefore, a much more precise determination of the (dimensionless) averaged spin can be reached in such measurements \cite{Yagi:2013bca}. 

In this paper we show that star-like solitonic solutions of different Skyrme-type models not only exist, but also reproduce to a good extent some of the currently best known properties of NS---like the typical values of mass, radius, moment of inertia, Love numbers, etc---coming from astrophysical measurements, GW observations and/or computer simulations.

We also address the issue of  whether the compact star solutions obtained within different Skyrme-based models and the corresponding EoS satisfy the I-Love-Q relations, and find that they indeed do satisfy them, even though the equations of state for different models present big differences.
The Skyrme model, being relatively simpler than other phenomenological or first-principle based relativistic field theories describing nuclear matter, therefore not only stands as an excellent candidate to describe nuclear matter at very high densities such as those inside NS. In addition, it provides a simple toolkit for the construction of a wide range of models of nuclear matter and their corresponding EoS which allow to investigate the resulting NS properties and universal relations in environments not considered previously---like, e.g.,  for non-barotropic EoS, see below.

The rest of the paper is organized as follows: In the second section, we present the generalized Skyrme model and the submodels that we want to consider, and find static solutions to the Einstein equations for these models coupled to gravity. For all models, we choose the model parameters such that the resulting skyrmionic matter approaches the nuclear saturation density $n_0=0.16 \, {\rm fm}^{-3}$ and the energy per baryon $E_0=923.3 \, {\rm MeV}$ of infinite nuclear matter at saturation in the limit of zero pressure, as in \cite{Adam:2020yfv}. In \cref{sec:Slowly-rot,Sec:Tid.demform}, we review the framework for obtaining perturbative solutions to the Einstein equations that represent, respectively, slowly rotating and tidally deformed stars. We find the equations for the metric perturbations up to second order in spin (for the rotating case) and to linear order in the external tidal field in the tidally deformed case, and solve these numerically for the star interior, after which the matching with the analytical exterior solutions is performed in order to obtain the correct values of the first multipoles in the asymptotic expansion of the metric outside the star. Also, in \cref{Sec:Tid.demform} the definition of the Love numbers is given and a procedure to obtain the first (quadrupolar) Love numbers from the perturbative analysis is explained.
Finally, in \cref{sect:Quasi-Univ-rel}, we show different quasi-universal relations between the dimensionless moment of inertia, quadrupolar moment, electric and magnetic Love numbers and compactness of the stars for all the models at hand, and discuss them. We conclude with a comment in \cref{sec:deform-constraints} about some constraints on the deformability of NS coming from recent GW observations of binary mergers and on the feasibility of the different Skyrme-based models to describe nuclear matter inside NS, and finish in the last section with a summary of all these results and prospects of future work. In our choice of units, the speed of light is $c=1$. For masses (energies) and lengths we use either solar masses $M_\odot$ and km---for astrophysical objects, or MeV and fm---for nuclear physics observables.

%%%%%%%%%%%%%%%%%%%%%%%%%%%%%%%%%%%
\section{Static stars and Generalized Skyrme model}
%%%%%%%%%%%%%%%%%%%%%%%%%%%%%%%%%%%
%%%%%%%%%%%%%%%%%%%%%%%%%%%%%%%%%%%
\subsection{The Generalized Skyrme model}
%%%%%%%%%%%%%%%%%%%%%%%%%%%%%%%%%%%
The original Skyrme model is defined by the following Lagrangian
\begin{align}
    \notag &\lag_{SK}\; =\; \lag_{2}+\lag_{4}+\lag_{0} \; =\;  -
    \frac{f_\pi^2}{4}\Tr{L_\mu L^\mu} \\ & \hspace*{0.8cm} +\frac{1}{32e^2}\Tr{[L_\mu,L_\nu][L^\mu,L^\nu]}-\mu^2\mathcal{U}(U),
    \label{Lsk}
\end{align}
where the Goldstone bosons associated to chiral symmetry breaking---the lightest degrees of freedom of QCD---form the $SU(2)$ matrix  Skyrme field $U(x)$.  Furthermore, $L_\mu=U^\dagger\partial_\mu U$ is the $\mathfrak{su}(2)-$valued, left invariant Maurer-Cartan form and $\mathcal{U}$ is a non-derivative part of the model, i.e., a potential. This theory possesses only three coupling constants $f_\pi,e$ and $\mu$ which are, respectively, the pion decay constant, the Skyrme coupling constant and the $\mu$ parameter related with the pion mass via $\mu=m_\pi f_\pi/\sqrt{8}$. More precisely, $m_\pi$ is the mass of small perturbations around the vacuum (pions) if the potential tends to the pion potential $\mathcal{U}_\pi=1/2\mbox{Tr} (1-U)$ for $U \to \mathbb{I}$. 

For finite energy solutions, it is necessary to impose constant boundary values of $U$ at $\abs{x}\rightarrow \infty$. This implies the appearance of a nontrivial topology. Indeed, the physically relevant matter field configurations define maps
\begin{equation}
    U:S^3\rightarrow SU(2)\simeq S^3,
    \label{maps}
\end{equation}
which are classified by an integer number or {\it topological degree}
\begin{equation}
    \mathcal{B}=\int B^0 d^3x, \hspace{2mm} B^{\mu} = \frac{1}{24\pi^2} \varepsilon^{\mu\nu\rho\sigma}\Tr{L_{\nu}L_{\rho}L_{\sigma}},
\end{equation}
where $B^{\mu}$ is the topological current. Importantly, it can be rigorously proven that the topological charge is just the baryon charge  \cite{Witten:1983tx}. Due to this equivalence, topological solitons with a non-zero value of the topological charge, typically referred to as Skyrmions, are identified as baryons and atomic nuclei, although the derivation of their properties as quantum systems from the Skyrme model (a non-renormalizable field theory) requires a careful quantization of the zero as well as massive (vibrational) modes of the classical solitons \cite{Adkins:1983ya,Halcrow:2016spb,Halcrow:2020gbm}. It is one of the most attractive features of the model that all these objects are emergent phenomena which arise from a very simple Lagrangian based entirely on pionic degrees of freedom, which contains a very small number of parameters. 

Being an effective theory, the Skyrme model can be extended by adding higher order terms to the Lagrangian. It can be shown that the only possible Lorentz-invariant extra term with at most second order time derivatives of the Skyrme field is \cite{Adam_2010}
\begin{equation}
    \lag_6=-\lambda^2\pi^4B_\mu B^\mu,
\end{equation}
where $\lambda$ is an additional coupling parameter related to the $\omega$-vector meson. Indeed, this term can be obtained by integrating out this vector meson from a model that includes both pions and vector mesons \cite{Vecmes2}. As a result, we get the {\it generalized Skyrme model}
\begin{equation}
\lag_{gen} = \lag_{SK} + \lag_{6}.
\end{equation}
For some choices of parameters and potentials \cite{Gudnason:2016tiz,Gudnason:2018jia}, this model maintains the successes of the standard Skyrme model (\ref{Lsk}) in the description of vibrational-rotational spectra of some light nuclei, but also leads to physical binding energies, which result too large in the standard Skyrme model.

An obvious, next step of application of the Skyrme model is to study properties of nuclear matter at extremely high densities, e.g., to describe the equation of state of neutron stars (see \cite{Naya_2019} for a recent review). To do so, we have to find the lowest energy solutions of the Skyrme model for a topological charge of the same order as the total baryon number of neutron stars, which typically is  $N \sim 10^{57}$. Basically, within the framework of the generalized Skyrme model, there are two qualitatively distinct possibilities. 

First of all, it is well known \cite{PhysRevD.40.3421,CASTILLEJO1989801} that the lowest energy solutions for the standard Skyrme model \eqref{Lsk} with an arbitrarily large baryon number, $B \to \infty$, consist of crystalline structures of Skyrmions. The EoS of this Skyrme crystal may then be used as a starting point for the investigation of NS. A second possibility is related to the fact that there is a very special point in the space of the model parameters, resulting in the so-called BPS Skyrme submodel:  
$
   \lag_{BPS} = \lag_6 + \lag_0 .
$
The name comes from the fact that this Skyrme model supports topological soliton configurations which saturate the BPS energy bound \cite{Adam_2010}, offering a possibility to resolve the problem of  the unphysically high binding energies of the standard Skyrme model. What is more important here, this model describes a perfect fluid for any value of the baryon charge. Indeed, the stress-energy tensor reads \cite{Adam_2015a,Adam_2015b}
\begin{equation}
    T_{BPS}^{\mu\nu}=(p\!+\!\rho)u^\mu u^\nu\!\!-\!p g^{\mu\nu},
    \label{BPSstress}
\end{equation}
 where the four-velocity $u^\mu$, pressure $p$ and energy density $\rho$ are
 \begin{equation}
      u^\mu=\frac{B^\mu}{\sqrt{B^\sigma B_\sigma}},\, p=\frac{\lambda^2\pi^4}{g}B^\mu B_\mu-\mu^2\mathcal{U},\,
    \rho = p + 2\mu^2\mathcal{U}.
    \label{BPSEoS}
 \end{equation}
 As the sextic term $ \lag_{6}$ gives the leading contribution to the energy at high pressure/density \cite{Adam:2015lra}, the fluid behavior is expected to dominate at this regime. This is, of course, consistent with the usual understanding of the inner core of neutron stars as being formed by a fluid of neutron matter. These two states of the Skyrmionic matter should be smoothly joined in the generalized model, suggesting a phase transition as pressure increases. Although the existence and properties of such a phase transition in the full model is still an unsolved problem, the known regimes at low and high pressure have recently led to the proposal of a {\it generalized EoS} \cite{Adam:2020yfv}, see below. 
 
%%%%%%%%%%%%%%%%%%%%%%%%%%%%%%%%%%%
\subsection{Static NS solutions}
%%%%%%%%%%%%%%%%%%%%%%%%%%%%%%%%%%%
Here and in the following sections, we will obtain solutions to the Einstein equations that describe NS within the different Skyrme models presented above. As a first step, we will consider static, spherically symmetric configurations, which is usually done following the Tolman-Oppenheimer-Volkoff (TOV) approach, in which the Einstein equations are solved using the stress-energy tensor of a perfect fluid. Thus, we suppose the spherically symmetric (Schwarzschild) ansatz for the metric,
\begin{equation}
    ds^2=-e^{\alpha(r)}dt^2+e^{\beta(r)}dr^2+r^2(d\theta^2+\sin^2\theta d\phi^2).
    \label{staticmetric}
\end{equation}
We extract from the Einstein equations
\begin{equation}
    R_{\mu\nu}-\frac{1}{2}Rg_{\mu\nu}=8\pi T_{\mu\nu}
    \label{EinsEqs}
\end{equation}
and the conservation of the stress-energy tensor of the perfect fluid type (${\nabla_\mu T^\mu_\nu=0}$) the following system of ODEs, also known as the TOV system,
\begin{subequations}
\label{TOV}
\begin{align}
    &\frac{d\alpha}{dr} = 2 \frac{4\pi r^3p + M}{r(r-2M)}\label{TOV1}\\[2mm]
    &\frac{dM}{dr} = 4\pi r^2 \rho \label{TOV2}\\[2mm]
    &\frac{dp}{dr} = -\frac{(p + \rho)}{2} \frac{d\alpha}{dr},\label{TOV3}
\end{align}
\end{subequations}
where we have made the usual definition 
\begin{equation}
    \exp(-\beta) = 1 - 2M/r
    \label{Mdef}
\end{equation}
so that the value $M_*=M(R_*)$ of the function $M = M(r)$ coincides with the (static) ADM mass of the star when evaluated at its radius $r=R_*$.

To close the system \eqref{TOV}, we have to know the relations between the pressure and the energy density, \textit{i.e.}, an EoS. It is at this point where the classical Skyrmion solutions with a very large value of the topological charge become relevant.

 %%%%%%%%%%%%%%%%%%%%%%%%%%%%%%%%%%%
\subsection{Skyrme neutron stars}
%%%%%%%%%%%%%%%%%%%%%%%%%%%%%%%%%%%
Next, we briefly review the current status of the description of static properties of neutron stars from the Skyrme model perspective.
%%%%%%%%%%%%%%%%%%%%%%%%%%%%%%%%%%%
\subsubsection{The $\lag_{SK}$ Skyrme neutron stars}
%%%%%%%%%%%%%%%%%%%%%%%%%%%%%%%%%%%
The usual Skyrme model is a field theory whose energy-momentum tensor does not have a 
perfect fluid form. Therefore, a suitable mean-field approximation has to be performed. In practice, it means a spatial averaging. The ground state is a crystal with a given lattice structure and lattice spacing $l_0$ (we assume the isotropic case). Obviously, the energy per baryon $E(l)$ has a minimum at $l=l_0$. This solution is also a zero-pressure (equilibrium) solution, because
\begin{equation}
p=-\frac{\partial E}{\partial V},
\end{equation}
where $V=l^3$ is the volume of the cell. Diminishing the lattice spacing $l$ is equivalent to imposing a nonzero pressure. Finally, as the pressure and the energy density are both functions of $l$, we can find the corresponding EoS, $\rho_{SK}=\rho_{SK}(p)$. If inserted into the TOV system, the crystal EoS amounts to neutron stars with rather small maximal masses, significantly below the observed NS masses. For example, for the cubic, face-centered lattice of $\mathcal{B}=4$ Skyrmions ($\alpha$ particles) $M_{\rm max}\simeq 1.49 M_\odot$ \cite{NelmesPiette,Naya_2019}. The corresponding mass-radius curve is presented in Fig. \ref{fig:MassvsRadius}, the pink dots.  
%%%%%%%%%%%%%%%%%%%%%%%%%%%%%%%%%%%
\subsubsection{The $\lag_{BPS}$ Skyrme neutron stars}
%%%%%%%%%%%%%%%%%%%%%%%%%%%%%%%%%%%
In the case of the BPS Skyrme submodel $\lag_{BPS}$, which is a genuine perfect fluid theory for any potential $\mathcal{U}$, one can find lowest energy Skyrmions for any value of the topological charge $\mathcal{B}$ in an exact form. There are, in fact, infinitely many solutions for a given $\mathcal{B}$ related via $SDiff$ diffeomorphisms, which corresponds very well with the fluid nature of the BPS Skyrmions. Interestingly, the perfect fluid form of the action allows to obtain the mean field EoS in an exact form without solving the field equations \cite{Adam_book,Adam_2015b}. This occurs because the pressure enters as an integration constant into the generalized Bogomolny equation. As a consequence, the pressure dependence of both the energy $E(p)$ and the volume $V(p)$ of BPS Skyrmions can be found as target space integrals (averages). The details of the resulting EoS obviously depend on the particular choice of the potential (but, of course, do not depend on a particular solution). On the other hand, since the sextic term provides the leading behavior in the high pressure limit, the EoS tends to the maximally stiff equation of state as the pressure increases
\begin{equation}
\rho_{BPS}(p) \approx p .
\end{equation}
\begin{figure}
    \hspace*{-0.45cm}
    \includegraphics[scale=0.45]{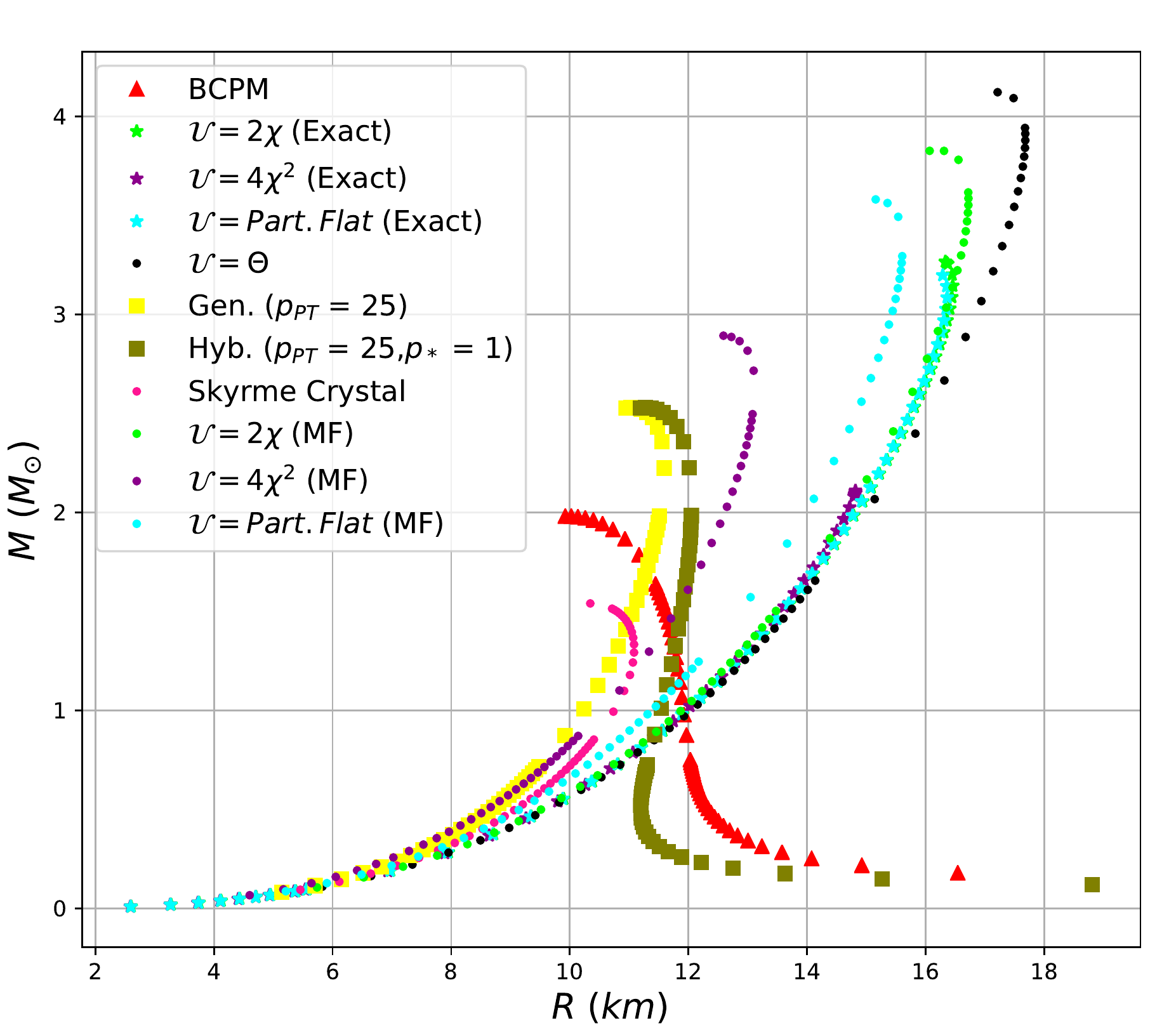}
    \caption{Mass-Radius diagram for the different EoS within the Skyrme model.}
    \label{fig:MassvsRadius}
\end{figure}
As a consequence of this stiffness, it is not surprising that the neutron stars provided by the BPS Skyrme model have rather big maximal masses, easily exceeding $3M_\odot$---see Fig. \ref{fig:MassvsRadius}, black, green, purple and blue dots, which correspond to the four different potentials introduced in \cite{Adam_book,Adam_2015b}, namely the $\theta$-potential $\mathcal{U}_\Theta=\Theta(\text{Tr}\{1-U\})$, the standard pion-mass potential $\mathcal{U}_\pi=1/2\text{Tr}\{1-U\}=2 \chi(r)$, the pion-mass squared potential $\mathcal{U}^2_\pi=4 \chi(r)^2$ and the partially flat potential \begin{equation}
    \mathcal{U}_{\rm PF}=\left\{\mqty{
    1,&\chi(r)\leq \tfrac{1}{2}\\\\
    16\chi^2(1-\chi)^2,&\chi(r)> \tfrac{1}{2}
    }\right..
\end{equation} 

Owing to its perfect fluid nature, the BPS model offers the possibility to close the TOV system without any mean-field approximation.  In this case, referred to as the exact case, the pressure and energy densities $\rho,p$  can already be read from the stress-energy tensor \eqref{BPSstress}. Furthermore, they are related in a non-algebraic way, by construction \eqref{BPSEoS}. This also means that the obtained matter is an example of a non-barotropic fluid where constant pressure do not correspond with constant energy density. Hence, this exact approach may serve as a laboratory where the impact of non-barotropic EoS on properties of NS can be studied. Further, the different BPS models provide  a wealth of new and different EoS which will allow us to test the universal, EoS-independent character of certain relations, like the \emph{I-Love-Q relations}, in new environments not considered previously.

More precisely, in the exact case the Skyrme field $U$ enters in the Einstein equations as an additional degree of freedom, so that we have to obtain its own differential equation in order to close the TOV system. To do this, we choose the hedgehog ansatz for the Skyrme field,
\begin{align}
    \notag&U(x)=e^{i\xi(r)\vu{n}(\theta,\phi)\cdot\,\bm{\sigma}},\\ &\vu{n}(\theta,\phi)=(\sin{\theta}\cos{(\mathcal{B}\phi)},\sin{\theta}\sin{(\mathcal{B}\phi)},\cos{\theta}),
\end{align}
which is compatible with the chosen ansatz of the metric, since it yields a spherically symmetric energy density, which is relevant for static NS. Here $\bm{\sigma}$ are the Pauli matrices and $(r,\theta, \phi)$ are spherical coordinates. The only degree of freedom in this ansatz corresponds to the radial profile $\xi(r)$, and inserting the hedgehog ansatz into the definition of $p$ it can be shown that this function satisfies the differential equation
\begin{equation}
    p = \frac{4\mathcal{B}^2\lambda^2 \xi'^2 \sin^4\xi}{e^{\beta} r^4} - \mu^2\mathcal{U},
\end{equation}
which is added to \eqref{TOV} to close the system.
For simplicity, when solving the TOV system we will define the new variable $\chi:=\sin^2(\xi/2)$, which satisfies 
\begin{equation}
    \frac{d\chi}{dr} = \frac{e^{\beta/2}r^2}{2\mathcal{B}\lambda}\sqrt{\frac{p+\mu^2\mathcal{U}}{\chi(1-\chi)}}.
    \label{hEq}
\end{equation}
Once the system of ODEs is closed, only a set of initial conditions are needed as an input in order to obtain a particular solution. 
However, in the exact case, the baryon number $\mathcal{B}$ of the star is an additional input parameter, and the value of the pressure at the center of the star ($p_0$) that yields the input value must be found via a shooting method,
with initial conditions
\begin{equation}
    \alpha(0) = \alpha_0, \hspace{1mm} M(0) = 0, \hspace{1mm} \chi(0) = 1, \hspace{1mm} p(0) = p_0,
\end{equation}
requiring that the pressure vanishes at some finite value $p(r = R_*) = 0$. This value $R_*$ is precisely the radius of the star. The value of $\alpha_0$ is not needed to solve  the system. However, only one value is correct, and it can be obtained by imposing continuity of the metric at the radius of the star, $R_*$, for which, and onwards, the metric is given by the Schwarzschild solution:
\begin{equation}
    e^{\alpha} = e^{-\beta} = 1 - \frac{2M_*}{r}, \quad r\geq R_*,\quad \text{where} \, M_*=M(R_*)
    \label{Matchingzerothorder}
\end{equation}
Also, the central value of the energy density  $\rho_0$ is determined by the BPS EoS \eqref{BPSEoS}. 

On the other hand, in the Skyrme crystal and the mean-field version of the BPS submodels, we do have a barotropic EoS $\rho(p)$, so that the energy density only depends on the pressure. In these cases, the equation \eqref{hEq} is no longer needed and the input parameter is the pressure in the center of the star $p_0$, along with the rest of initial conditions for $\alpha$ and $M$. The system of differential equations is then solved up to the star radius ($R_*$), in that point the static ADM mass of the star $M_* = M(R_*)$ is also obtained. 

In Fig. \ref{fig:MassvsRadius} mass-radius curves for the exact case are presented - see green, blue and purple stars. For the $\theta$-potential the MF and exact computations obviously coincide. Therefore, for relatively flat potentials (e.g., the pion-mass and the partially flat potential) the difference between the MF and exact approach is rather small, while it strongly increases for more peaked  potentials (e.g., the pion-mass potential squared).
%%%%%%%%%%%%%%%%%%%%%%%%%%%%%%%%%%%
\subsubsection{Neutron stars and the generalized EoS}
%%%%%%%%%%%%%%%%%%%%%%%%%%%%%%%%%%%
As we see, the usual Skyrme model crystal and the BPS Skyrme fluid result in too small or too large maximal masses of neutron stars, respectively. It can be expected that these two extremal cases can be balanced in the full generalized Skyrme model. While the EoS for the generalized Skyrme model is not currently available, it motivates the following generalized Skyrme EoS which interpolates between the crystal and fluid phases \cite{Adam:2020yfv}
\begin{equation}
  \rho_{\rm Gen}(p)=(1-\alpha(p)) \rho_{SK}+\alpha(p) (p+\rho_{SK} (p_{PT}))
\end{equation}
where the interpolating function 
\begin{equation} \label{alpha}
 \alpha(p,p_{PT}, \beta) = \frac{\left( \frac{p}{p_{PT}} \right)^\beta}{1+\left( \frac{p}{p_{PT}}\right)^\beta}
\end{equation}
tends from 0 for $p/p_{PT} \to 0$ to 1 for $p/p_{PT} \to \infty$. The parameter $p_{PT}$ can be identified with the position of the crystal/fluid phase transition, whereas $\beta$ measures how rapid the transition occurs. Specifically, we assumed a rather gradual phase transition ($\beta=0.9$) located at $p_{PT} \in (25,50) \mbox{MeV/fm}^3$ in \cite{Adam:2020yfv}. We remark that the value of $p_{PT}$ strongly affects the maximal mass. 

In Fig. \ref{fig:MassvsRadius} we show the mass-radius curve for the generalized EoS with $p_{PT}=25 \mbox{MeV/fm3}^3$ - see yellow squares. As expected, the maximal mass of NS is between the two previously discussed versions of the Skyrme model and reads $M_{\rm max} \simeq 2.55 M_\odot$.
%%%%%%%%%%%%%%%%%%%%%%%%%%%%%%%%%%%
\subsubsection{Neutron star crusts and the hybrid EoS}
%%%%%%%%%%%%%%%%%%%%%%%%%%%%%%%%%%%
By construction, the generalized Skyrme model contains only pionic degrees of freedom (with some other heavier mesons effectively also taken into account). This means that it is relevant for describing nuclear matter above the saturation density. For lower densities, the electromagnetic interaction starts to have a nontrivial impact on the properties of nuclear matter, leading to the appearance of inhomogeneous phases (such as "nuclear pasta" phases \cite{Chamel_2008}). Although the Skyrme model can be coupled with the electromagnetic $U(1)$ gauge field, which in principle may allow to study such phases within the framework of the Skyrme model, the resulting theory is very complicated and currently no large $B$ Skyrmions are known. However, it is possible to take into account this low density regime, relevant for the crust region of NS, by assuming a transition of the generalized Skyrme EoS to a standard nuclear EoS obtained by the usual many-body techniques. Concretely, we choose the EoS $\rho_{\rm BCPM}$ of \cite{Vinhas_2015}, as we did in \cite{Adam:2020yfv}. As a consequence, we arrive at a {\it hybrid} EoS 
\begin{equation} \label{hyb}
  \rho_{\rm Hyb}(p)=(1-\alpha(p)) \rho_{\rm BCPM}+\alpha(p) (p+\rho_{\rm Gen} (p_*))
\end{equation}
where now $\beta=2$ and the position of the transition $p_* \in [0.5,2] \mbox{MeV/fm}^3.$ Further, $\alpha (p,p_*, \beta)$ is defined in (\ref{alpha}). The resulting mass-radius curve is presented in Fig. \ref{fig:MassvsRadius}, olive squares (for $p_*=1\, {\rm MeV}\,{\rm fm}^{-3} $ and $p_{PT}=25\, {\rm MeV}\,{\rm fm}^{-3} $). We want to emphasize that the NS resulting from the hybrid EoS (\ref{hyb}) pass {\it all} current observational constraints.

%%%%%%%%%%%%%%%%%%%%%%%%%%%%%%%%%%%
\section{Slowly rotating Skyrme stars }
%%%%%%%%%%%%%%%%%%%%%%%%%%%%%%%%%%%
\label{sec:Slowly-rot}
In this section, we will study how the previously obtained spherically symmetric Skyrmion stars---models of NS based on the (generalized) Skyrme models considered in the present paper---behave under small perturbations. In particular, we will analyze their deformation due to (small) rotation and tidal forces, in order to obtain useful relations between their moment of inertia, deformability and Love number---known as \emph{I-Love-Q} relations, first proposed in \cite{Yagi_2013}---which may help in extracting information about the internal structure of compact stars. Throughout this and subsequent sections, we will largely follow  the approach and notation of \cite{Yagi_2013}.
%%%%%%%%%%%%%%%%
\subsection{Slowly rotating stars: Hartle-Thorne formalism}
\label{sect:HartleThorne}
%%%%%%%%%%%%%%%
 To analyze the properties of rotating Skyrmion stars, we will make use of the Hartle-Thorne formalism for slowly rotating stars, introduced in \cite{HartleI}. This formalism establishes a perturbative framework which consists in an expansion of the metric in powers of a perturbation parameter---related with the rotational frequency---and solving the Einstein equations order by order in this parameter. 
This perturbative expansion has proven particularly useful in the literature since it allows to obtain approximate solutions to the Einstein equations both for the interior and exterior of the star, hence, it enables to retrieve information about the equation of state for the matter inside the star from the multipolar expansion of the external solution. We will now review the procedure to obtain the solution in this approximation for the metric in the interior of a compact star, and in the following sections we will do the same for the exterior solution and the matching between both solutions at the star surface.

The starting point of the slow rotation approximation is to consider a static solution for the metric of a non-rotating configuration, and subsequently add perturbation terms up to a given order in a suitable parameter related to the spin of the star. In our case, we will start from the static metric with line element \eqref{staticmetric} %\begin{equation}
    %ds^2= -e^{\alpha(r)}dt^2+e^{\beta(r)}dr^2+r^2(d\theta^2+\sin^2\theta d\phi^2),
    %\label{staticmetric}
%\end{equation}
and, as in  \cite{Berti2005}, defining the spin parameter $\epsilon=\Omega_*/\Omega_K$
in terms of $\Omega_*$---the angular velocity of the star as measured by an external, static observer located at spatial infinity---and the characteristic frequency 
$
    \Omega_K=\sqrt{M_0/R^3_0}
$, 
where $M_0$ and $R_0$ are the non-spinning mass
and radius of the star. The characteristic frequency
$\Omega_k$ corresponds to the Keplerian orbital period of
a test particle at a radius $R_0$ around a mass $M_0$ and thus
can be thought of as the rotational frequency for which the mass shedding occurs, i.e., an upper limit for the rotational frequency of the star \cite{Berti_2013}. For spin frequencies
much smaller than this characteristic frequency, the parameter $\epsilon$ serves as a suitable small parameter about
which we can expand the metric. On the other hand, for spin frequencies near the Keplerian limit, $\epsilon\sim 1$ and the Hartle-Thorne approximation is no longer valid. Despite the dependence of the Keplerian frequency on the EoS, the slow-rotation approximation is valid
for even the most rapidly spinning neutron stars observed
to date \cite{Berti2005}.

Therefore, let us consider the background spacetime whose metric is given by the static line element \eqref{staticmetric}. We now extend this metric by defining a one-parameter family of metrics $g(\epsilon)$  whose components may be expanded in powers of $\epsilon$, $g(\epsilon)=g^{(0)}+\epsilon g^{(1)}+\tfrac 1 2 \epsilon^2 g^{(2)}+ \cdots$, with $g^{(0)}$ given by \eqref{staticmetric}. Note that this construction introduces an inherent gauge freedom (for details see, for example,  \cite{Sopuerta_2004,Reina_2015}) Thereby, following \cite{Yagi_2013}, up to second order in $\epsilon$, we may write the metric of a slowly rotating star in the Regge-Wheeler gauge as:
\begin{align}
    ds^2=-(1&+2\epsilon^2\bar{h})e^{\bar\alpha}dt^2+\qty(1+2\epsilon^2\frac{\bar{m}}{r-2\bar M})e^{\bar\beta}dr^2+\nonumber\\
    +&(1+2\epsilon^2\bar{k})r^2\qty[d\theta^2+\sin^2(\theta)(d\phi+\epsilon\bar{\omega}dt)^2],
    \label{perturbmetric}
\end{align}
 where $\bar\omega=\bar \omega(\theta,r)$, $\bar h=\bar h(\theta,r)$, $\bar m=\bar m(\theta,r)$, $\bar k=\bar k(\theta,r)$, and $\bar M(r)$ is related to $\bar \beta(r)$ in the same form as in \eqref{Mdef}.  Comparing with the general expansion of $g(\epsilon)$, we find:
\begin{align}
    g^{(1)}&=2r^2\bar{\omega}\sin^2\theta dtd\phi,\\g^{(2)}&=-\left(4e^{\bar\alpha}\bar{h}+2r^2\sin^2\theta\bar{\omega}^2\right)dt^2+\nonumber\\
    &\quad+4e^{\bar\beta}\frac{\bar{m}}{r-2\bar{M}}dr^2+4r^2\bar{k}(d\theta^2+\sin^2\theta d\phi^2).
\end{align}

Note that the metric perturbation function $\bar\omega$ enters at first order in the spin parameter, whereas $\bar{h}$, $\bar{m}$ and $\bar{k}$ correspond to second order perturbations. This can be easily understood with the following argument \cite{HartleI}: a transformation of the metric for a stationary and axially symmetric rotating spacetime of the form $\Omega\rightarrow-\Omega$ should be equivalent to $t\rightarrow -t$. 
This, in particular, implies that an expansion of the diagonal components of the metric in powers of $\epsilon$ must contain only even powers (since they are unchanged under time reversal), whilst an expansion of the $g_{0,3}$ term will only contain odd powers of $\epsilon$. Furthermore, since $\bar{\omega}$ corresponds essentially with the $g_{03}$ term of the metric, it is responsible for the dragging of inertial frames. In other words, it measures the rate of rotation that a freely falling observer would undergo with respect to a static one (Lense-Thirring effect).

Due to these perturbation terms in the spacetime metric, both the Einstein tensor for the metric and the stress-energy tensor for the matter field will develop perturbation terms, as well. Indeed, just as with the metric tensor, we may define the one-parameter families of perturbed quantities $G_{\mu\nu}(\eps)$ and $T_{\mu\nu}(\eps)$, expand them in powers of $\eps$ and impose that Einstein equations are satisfied order by order in the expansion parameter. In particular, both the pressure and mass densities of the matter field will be perturbed, acquiring an angular dependence, i.e.
\begin{align}
    p(\eps;r,\theta)&=p_0(r)+\epsilon p_1(r,\theta)+\frac{1}{2}\epsilon^2p_2(r,\theta)+\mathcal{O}(\eps^3),\\ \rho(\eps;r,\theta)&=\rho_0(r)+\epsilon\rho_1(r,\theta)+\frac{1}{2}\epsilon^2\rho_2(r,\theta)+\mathcal{O}(\eps^3),
    \label{pressperturb}
\end{align}
as well as the fluid four-velocity, $u(\eps)$. For this latter quantity, we further impose the normalization condition $g(\eps)_{\mu\nu}u^\mu(\eps)u^\nu(\eps)=1$. Also, stationarity, axial symmetry and rigidity of the fluid flow requires $u(\eps)$ to be proportional to both killing vectors, i.e. $u(\eps)=f_1(\eps)(\partial_t+f_2(\eps)\partial_\phi)$.  The $f_1$ function is obtained by the normalization condition at each order, and, since the background configuration corresponds to a static fluid, $f_2(\eps)=\eps C+\mathcal{O}(\eps^3)$. We therefore have 
\begin{equation}
    u(\eps)^\mu=(u^t(\eps),0,0,\eps C u^t(\eps)),
    \label{nogauge4velocity}
\end{equation}
thus the constant $C$ corresponds to the angular velocity of the fluid as measured within the inner coordinate system.
Note also that only odd powers of $\eps$ enter the expansion of $f_2$, for the same symmetry arguments as for $\bar{\omega}$.

It is important to notice that all these (one-parameter families of) objects so defined are gauge-dependent, although the Einstein equations themselves do not depend on the gauge (i.e, they must be fulfilled in any gauge). We thus may take advantage of this gauge freedom to choose the most convenient form of the metric functions. In particular, we may choose $C=\Omega_K$ in \eqref{nogauge4velocity}, so that the coordinate system in the interior of the star is taken to be that of a static observer which measures the angular velocity of the fluid to be $d u^\phi/d u^t=\varepsilon\Omega_K=\Omega_*$. It can be shown that any other choice of the constant $C=C_0$ is equivalent to a gauge transformation of the first order metric perturbation defined by the vector $V=(\Omega_K-C_0)t\partial_\phi $\cite{Reina_2015}.

On the other hand, the coordinate system we have chosen so far is not quite well suited to perform the integration of the Einstein field equations from the inside of the star, for the following reason: in order to find a numerical solution for the interior metric, we will have to solve Einstein equations with a non-vanishing stress-energy tensor up to the surface of the star, which is usually defined by the surface of vanishing pressure. While in the spherically symmetric case the surfaces of constant density (or pressure) are trivially those of constant radial coordinate, this is no longer the case once the second order perturbations of the metric due to rotation are taken into account.  Indeed, the (perturbed) pressure and mass densities \eqref{pressperturb} will depend both on $r$ and $\theta$, so that the surface of the star will be deformed with respect to the static case. 

Therefore, we will consider a choice of gauge in which the surfaces of constant pressure (density) of the perturbed configuration are those of constant radial coordinate. This is in fact equivalent to a change of coordinates in the perturbed configuration from the original (background) coordinate system $\{t,r,\theta,\phi\}$ to another, $\{t,\bar{r},\theta,\phi\}$, in which the new radial coordinate is defined by
\begin{equation}
    p(\eps;r,\theta)=p_0(\bar{r}),\quad r\equiv r(\eps;\bar{r},\theta)=\bar{r}+\eps^2\zeta(\bar{r},\theta)+\mathcal{O}(\eps^3)
\end{equation}
so that $r$ coincides with $\bar{r}$ in the background configuration ($\eps=0$), whilst the function $\zeta(\bar{r},\theta)$ measures the deviation from spherical symmetry of the perturbed configurations. 
The new radial coordinate $\bar{r}$ is defined so that $p_0(\bar r)=\text{const}$ defines the isobaric surfaces of the rotating star. 

Strictly speaking, one could think that, in the exact case, also the perturbations of the Skyrme profile function $\chi$ must be taken into account. However, these will be by construction directly related to the energy and pressure perturbations, and, since we will get rid of these perturbations by a suitable radial coordinate change, also the perturbation on the radial Skyrme profile will disappear. We have checked that this is in fact the case, and that no extra degrees of freedom appear in the perturbative formalism for the exact BPS Skyrme case up to second order in $\epsilon$.

In the new coordinate system, the metric \eqref{perturbmetric} is rewritten, up to second order in $\eps$:
\begin{equation}
\begin{split}
 ds^2=&-(1+2\epsilon^2\bar{h})\qty(1+\eps^2\bar \alpha'\zeta)e^{\bar{\alpha}}dt^2+2\eps^2e^{\bar{\beta}}\partial_{\theta}\zeta d\rb d\theta+\\+&\qty[1+2\epsilon^2\qty(\frac{\bar{m}}{\rb-2\bar M}+\partial_{\rb}\zeta)](1+\eps^2\bar \beta'\zeta)e^{\bar{\beta}}d\rb^2+\\
 +&(1+2\epsilon^2\bar{k})(\rb^2+2\eps^2\rb\zeta)\qty[d\theta^2+\sin^2\theta(d\phi+\epsilon\bar{\omega}dt)^2],
\end{split} \label{uglymetric}
\end{equation}
where all the metric functions are written as functions of $\rb$ (and possibly $\theta$), and the $'$ denotes a derivative with respect to $\rb$.

The metric \eqref{uglymetric} has a rather complicated form. However, we may simplify it by redefining, the metric functions:
\begin{equation}
\begin{array}{clcl}
   e^ \alpha(\rb)=&e^{\bar{\alpha}(r)}=e^{\bar{\alpha}(\rb)}(1+\eps^2\bar{\alpha}'(\rb)\zeta(\rb,\theta)+\mathcal{O}(\eps^3)), \\
    e^\beta(\rb)=&e^{\bar{\beta}(r)}=e^{\bar{\beta}(\rb)}(1+\eps^2\bar{\beta}'(\rb)\zeta(\rb,\theta)+\mathcal{O}(\eps^3)),\\
    \varpi(\rb,\theta)=&\bar{\omega}(\rb,\theta),\quad M(\rb)=\bar M(\rb),\\
    k(\rb,\theta)=&\bar{k}(\rb,\theta),\, m(\rb,\theta)=\bar{m}(\rb,\theta),\, h(\rb,\theta)=\bar{h}(\rb,\theta)
\end{array}
\end{equation}
so that the new metric
\begin{equation}
\begin{split}
 ds^2&=-(1+2\epsilon^2{h})e^{{\alpha}}dt^2+\\
 &+\qty[1+2\epsilon^2\qty(\frac{{m}}{\rb-2 M}+\partial_{\rb}\zeta)]e^{{\beta}}d\rb^2+2\eps^2e^{{\beta}}\partial_{\theta}\zeta d\rb d\theta+\\
 &+(\rb^2+2\epsilon^2\rb^2({k}+\zeta/\rb))\qty[d\theta^2+\sin^2\theta(d\phi+\epsilon{\varpi}dt)^2],
\end{split} \label{nicemetric}
\end{equation}
coincides with \eqref{uglymetric} up to second order in $\epsilon$. Although both metrics \eqref{nicemetric} and \eqref{perturbmetric} are different, they are related through a gauge transformation, so that both must satisfy Einstein equations, and the gauge-independent results obtained in both approaches must be the same (at least, up to second order in $\eps$). Note that these metrics are compatible with the general form for the Hartle-Thorne metric in an arbitrary gauge, obtained in \cite{Reina_2015}, which have two commuting killing vector fields $k_{(\phi)}=\partial_\phi$ and $k_{(t)}=\partial_t$.

Although a priori the metric perturbation functions can have an arbitrary dependence on $r $ and $\theta$, an expansion of these functions is always possible in spherical harmonics \cite{HartleI}. Moreover, the angular dependence of the perturbation functions may be further reduced by additional arguments. For example, axial and reflection symmetry in the equatorial plane implies that the $m$ (axial) number in the spherical harmonic expansion does not play any role, so that it may be reduced to an expansion in terms of Legendre polynomials. Therefore, we may expand the metric perturbation functions into a series of Legendre polynomials or their derivatives, depending on the parity of the corresponding perturbation function (see eg  \cite{Martel_2005} or \cref{Appendperturb} for details). Thus, for the odd parity perturbation function $\varpi(\rb,\theta)$, we have
\begin{equation}
    \varpi(\rb,\theta)=\sum_l\varpi_l(\rb)\dv{}{\cos\theta}P_l(\cos\theta),
    \label{omegl}
\end{equation}
whereas for the even parity functions $h$, $m$ and $k$:
\begin{equation}
\begin{aligned}
    h(\rb,\theta)&=\sum_l h_l(\rb)P_l(\cos\theta),\\ m(\rb,\theta)&=\sum_l m_l(\rb)P_l(\cos\theta),\\ k(\rb,\theta)&=\sum_l k_l(\rb)P_l(\cos\theta),
\end{aligned}
\end{equation}
and the same holds for $\zeta(\rb,\theta)$. Furthermore, one can show that the requirements of asymptotic flatness  and regularity of the metric at the center of the star impose that only the $l=0$ term of \eqref{omegl} survives, and similar arguments can be made for the second order perturbation functions, in which case only the $l=0,2$ terms are non vanishing \cite{HartleI}. Thus, the spacetime metric is reduced to \eqref{nicemetric}, where $\varpi(\rb,\theta) =\varpi_1(\rb)\equiv \varpi(\rb)$, $    h(\rb,\theta)=h_0(\rb)+h_2(\rb)P_2(\cos\theta) 
$, and so forth.  Also, in the following, it will become useful to work with the shifted function $\omega(\rb)$ defined by 
\begin{equation}
\varpi(\rb)\equiv (\Omega_K-\omega(\rb)).
\end{equation}
 Furthermore, we can make use of the residual gauge freedom of reparametrizations of the radial coordinate to set $k_0(\rb)=0$ in the expansion.

On the other hand, with these gauge choices, the stress-energy tensor of the system will be given by
\begin{equation}
    T^\mu_\nu=(\rho(\bar{r})+p(\bar{r}))u^\mu u_\nu+p(\bar{r})\delta^{\mu}_\nu,
\end{equation}
where, from the normalization condition for the four-velocity, we have $u^\mu=u^t(1,0,0,\Omega_*)$, and
\begin{equation}
    u^t=\frac{1}{\sqrt{-g_{tt}-2\Omega_*g_{t\phi}-\Omega_*^2g_{\phi\phi}}},
\end{equation}
which, up to second order in $\epsilon$, reads

\begin{equation}
    u^t=e^{-\frac{\alpha}{2}}+\epsilon^{2}{\qty(\frac{\rb^{2}}{2} \omega^{2} \sin^2{\theta}- [h_0+h_2P_2(\cos\theta)] e^{\alpha})}  e^{-\frac{3\alpha}{2}}.
\end{equation}
 
To sum up, we have described the metric of spacetime associated to a slowly rotating perfect fluid star up to second order in the spin parameter. To do so, a perturbative expansion must be performed from a spherically symmetric, non-rotating metric in terms of a certain set of perturbation functions. We have chosen a particular coordinate system in which the surfaces of constant pressure coincide with those of constant radial coordinate, and written the stress-energy tensor of the rotating fluid in these coordinates. Therefore, we are now ready to obtain the Einstein equations for the system.
\subsubsection{Interior Einstein equations}
Thus, we now consider the Einstein equations \eqref{EinsEqs}, which can be written as $E = 0$ with $E := G - 8\pi T$, for the interior metric  \eqref{nicemetric}.
The Einstein equations imply different equations for the perturbation functions, at each order in $\epsilon$. Indeed, we may write 
\begin{equation}
    E(\eps)=E^{(0)}+E^{(1)}\epsilon+\frac{1}{2}E^{(2)}\epsilon^2+\cdots
    \label{PertEinstein}
\end{equation}
where $E^{(1)}=\partial_\epsilon E|_{\epsilon=0}$, and so forth, so that all terms in this expansion must vanish.
For example, the zeroth-order equations correspond to the TOV system of equations \eqref{TOV}. 

At first order in $\eps$, the only nontrivial equation is obtained from the first-order Einstein equation $E^{(1)}=0$, and corresponds to the $(t,\phi)$ component, $    {E^{(1)\phi}}_t=0$, which yields
\begin{equation}
\omega'' =  4{\left(\pi \rb e^{\beta} (p + \rho) -\frac  1 \rb\right)}\omega'+{16 \, \pi e^{\beta}  (p +\rho) }\omega.
\label{omedott}
\end{equation}

The second order Einstein equations are given by $E^{(2)}=0$. As we have seen, the second order perturbation functions can be divided into two sectors, corresponding to the $l=0$ and $l=2$ terms in the Legendre expansion. Furthermore, these sectors appear uncoupled in the Einstein equations, so that we may separate these into different sets of equations for each sector. At quadratic order in $\eps$, it will also be useful to consider, apart from the Einstein equations, the stress-energy tensor conservation equation. In particular,  from the $l=0$ sector of $\nabla_\mu T^{(2)\mu}_{\quad r}=0$, one finds
\begin{equation}
    h_0'=\frac 13 \qty(\rb^2\omega^2 e^{-\beta})'-\frac 12  \qty[\frac{\zeta_0}{\rb}e^{\alpha}(2M+8\pi \rb^2p)]'.
    \label{h0zeta0}
\end{equation}
This equation can be integrated to yield an algebraic equation for $\zeta_0$ in terms of $h_0$ and its initial condition, $h_0^{(0)}$ which is in a priori unknown and will be determined once the system is solved by matching with the exterior solutions. Further, from $\nabla_\mu T^{(2)\mu}_{\quad \theta}=0$, we have
\begin{equation}
    \zeta_{2} = %-\frac{2 \, \left(\rb^{2} \omega^{2} + 3 \, e^{\alpha} h_{2} \right)e^{-\alpha}}{3 \, \alpha'}=
    -\frac{{\left(\rb^{2} - 2 \, \rb M\right)}\qty[\rb^2 e^{-\alpha} \omega^{2} + 3 \, h_{2}]}{3 \, {\left(4 \, \pi \rb^{3}  p + M \right)}},
    \label{zetadot}
\end{equation}
where we have used the zeroth order TOV equations. The last two equations are only valid inside the star since we are supposing $p, \rho \neq 0$. In particular, as they both correspond to algebraic instead of differential equations, the variable $\zeta_2$ will not appear in the second-order system of differential equations since we can substitute directly by \eqref{zetadot}, and the same will happen to $h_0'$.

Let us now obtain the differential equations for the rest of metric perturbation functions. The $l=0$ contribution of ${E^{(2)}}^t_t=0$ gives
\begin{equation}
    m_{0}' = \frac{8}{3}\pi \rb^{4} e^{-\alpha} \omega^{2} (\rho+p) + \frac{1}{12} \, \rb^{4} e^{-\left(\alpha+\beta\right)} (\omega')^{2} - 4 \, \pi \rb^{2} \zeta_{0}\rho'.
    \label{m0eq}
\end{equation}
For the $l=2 $ sector, ${E^{(2)\theta}_{\quad\theta}-E^{(2)\phi}_{\quad\phi}=0}$ and  $E^{(2)r}_{\quad\theta}=0$ yield, respectively,
\begin{align}
    m_{2} \!&= \!\! \, {\left[\frac{8}{3} \, \pi \rb^{5} e^{\beta} \omega^{2}( p +\rho) + \frac{\rb^{5}}{6}(\omega')^{2} - \, \rb e^{\left(\alpha+\beta\right)} h_{2}\right]} e^{-\left(\alpha+2 \beta\right)},\label{mdos}\\
    k_2'&=-h_2'+\frac{\rb-3 M-4 \pi p \rb^{3}}{\rb^{2}} e^{\beta} h_{2}+\frac{\rb-M+4 \pi p \rb^{3}}{\rb^{3}} e^{2 \beta} m_{2}.\label{kdot}
\end{align}
On the other hand, the Einstein equation $E^{(2)r}_{\quad r}=0$ yields two independent equations which must be satisfied separately, namely, one for the $l=2$ sector (obtained from the terms proportional to $P_2(\cos\theta)$), and other for the $l=0$ sector (from the terms independent of the Legendre polynomial). The equation for the $l=2$ sector can be written 
\begin{equation}
    \begin{aligned}
h_2'=&-\frac{\rb-M+4 \pi p \rb^{3}}{\rb} e^{\beta} k_2'+\frac{3-4 \pi(\rho+p) \rb^{2}}{\rb} e^{\beta} h_{2}+\\&+\frac{2}{\rb} e^{\beta} k_{2}+\frac{1+8 \pi p \rb^{2}}{\rb^{2}} e^{2 \beta} m_{2}+\frac{\rb^{3}}{12} e^{-\alpha}\left(\omega'\right)^{2} 
-\\&-\frac{4 \pi(\rho+p) \rb^{4} \omega_{1}^{2}}{3 \rb} e^{-\alpha+\beta},
\label{hdot}
\end{aligned}
\end{equation}

whereas for the $l=0$ sector we have
\begin{equation}
\begin{split}
    h_0'=-&\frac{r^3}{12}(\omega')^2 e^{-\alpha}-\frac{e^{\beta}}{2\rb^2}\qty(\frac{2M}{\rb}-\rb\alpha')\zeta_0-\frac{\alpha''}{2}\zeta_0\\
    +&\qty[\frac{m_0}{\rb^2}+\qty(4\pi \rho-\frac{2M}{\rb^3})\zeta_0]e^{\beta}(1+\rb\alpha').
    \label{h0eq}
\end{split}
\end{equation}
Substitution of \cref{h0zeta0} into \eqref{h0eq} yields a differential equation for $\zeta_0$, which, together with \cref{m0eq}, constitutes a system of two ODEs independent of $h_0$. Thus, once this system is solved, $h_0$ can be found algebraically using the integrated version of \eqref{h0zeta0} up to an arbitrary constant.

\subsubsection{Exterior equations and solutions}
Following \cite{HartleI,Yagi_2013}, we may take \eqref{perturbmetric} as an ansatz for the metric of spacetime in the star exterior. We can indeed do this, since $\zeta(r,\theta)$ is defined only inside the star, and taken to be constant outside. This means that the exterior metric in terms of $r$ and $\rb$ will be the same, where now the radial coordinate $r$ goes from a finite value in the star surface $R_*$, --corresponding to the star radius at zeroth-order-- to infinity. Following the same steps as in the previous section, the Einstein equations in the exterior of the star can be obtained at each order in $\eps$ simply by setting $\rho=p=0$ in \Cref{omedott,mdos,kdot,hdot}, with the exterior solution of the zeroth-order equations (TOV system) corresponding to the Schwarzschild solution by virtue of Birkhoff's theorem.
Hence, to first order in $\eps$, we have:
$
   \eps \omega^{\text {ext }}=K_1-{K_2/r^{3}}
$
where $K_1$ and $K_2$ are two integration constants which can be related to the total spin velocity and angular momentum of the star. Indeed, at $r\rightarrow\infty$, the metric function $\eps\omega^{\text{ext}}$ must approach the angular velocity of the star as measured by an static observer,  so that ${K_1=\Omega_*}$. On the other hand, we may calculate the conserved total angular momentum $J$ of the star by integrating the angular momentum density current $J^\mu={T^\mu}_\nu k_{(\phi)}^\nu$ over a spacelike hypersurface $\Sigma$:
\begin{equation}
    J=\int_{\Sigma} T_{\mu \nu} k^{\mu} dS_\mu=\int T_{\varphi}^{t}|g|^{1 / 2} d r d \theta d \phi = K_2/2+\order{\eps^2}
\end{equation}
from where it is straightforward to see that 
\begin{equation}
   \eps \omega^{ext}=\Omega_*-\frac{2J}{r^3}.
   \label{omegaext}
\end{equation}
On the other hand, at second order in $\eps$, the system given by \cref{mdos,kdot,hdot} must be solved with vanishing $ \rho $ and $p$. Using the expressions for the exterior solution of the first and zeroth-order metric functions, and imposing asymptotic flatness of the metric, one finds the analytic expressions \cite{HartleI,Yagi_2013}:
\begin{widetext}
\begin{align}
 m_{0}^{\text {ext }} &=\delta M-\frac{J^2}{r^3}, \\
 h_{0}^{\text {ext }} &=-\frac{\delta M}{(r-2M)}+\frac{J^2}{r^3(r-2M)}, \\
    h_{2}^{\text {ext }} &=\frac{1}{M_{*}r^{3}}\left(1+\frac{M_{*}}{r}\right) J^{2}+A Q_{2}^{2}\left(\frac{r}{M_{*}}-1\right), \\
    k_{2}^{\text {ext }} &=-\frac{1}{M_{*} r^{3}}\left(1+\frac{2 M_{*}}{r}\right) J^{2}+\frac{2 A M_{*}}{\sqrt{r\left(r-2 M_{*}\right)}} Q_{2}^{1}\left(\frac{r}{M_{*}}-1\right)-A Q_{2}^{2}\left(\frac{r}{M_{*}}-1\right), \\
    m_{2}^{\text {ext }} &=-\frac{1}{M_{*} r^{2}}\left(1-7 \frac{M_{*}}{r}+10 \frac{M_{*}^{2}}{r^{2}}\right) J^{2}+\frac{3 A r^{2}}{M_{*}}\left[1-3 \frac{M_{*}}{r}+\frac{4}{3} \frac{M_{*}^{2}}{r^{2}}+\frac{2}{3} \frac{M_{*}^{3}}{r^{3}}+\frac{r}{2 M_{*}} f(r)^{2} \ln f(r)\right],
\label{ExteriorSol}
\end{align}
\end{widetext}
being $f(r) = \left( 1-2M_*/r \right)$,  and where $\delta M$ and $A$ are integration constants. As we will see, $\delta M$ corresponds to the correction of the gravitational mass, whereas $A$ will be associated to the Love numbers.

\subsubsection{Numerical
solution for the interior and matching}
Once we have obtained the system of differential equations for the metric functions, we need now the initial conditions in order to solve it. In this section we will explain how to obtain them and also how to solve the system numerically.

At this point there are no differences between how to solve the exact case and the mean-field case since the shooting method for the exact case is required only for the zeroth-order equations and those have already been solved. Thus we do know which value of the pressure in the center of the star corresponds to a given baryon number.

To obtain the initial conditions, as before, we expand our metric functions in powers of the radial coordinate and insert them in the differential equations to obtain the relations between the coefficients. In the zeroth-order (non-rotating) problem it is enough to expand until the zeroth order coefficient (in powers of $\rb$), however when dealing with the second-order functions we need to expand them to the first nontrivial order (with nonvanishing coefficients). The reason is that the metric functions $h_2$ and $k_2$ vanish at the center of the star. Furthermore, for the next term of the expansions we find that they are equal and opposite, thus cancelling each other when substituting into their equations. This implies that: 1. We need a really good accuracy in the step of the numerical integration and 2. We cannot obtain the value of the first nontrivial coefficient of $h_2$, in its expansion in powers of $\rb$.  To solve both problems, we follow \cite{Yagi:2013mbt} and start the integration in some small radius $R_{\epsilon}$ (instead of $\rb=0$) such that the expansions \eqref{InitialConditions} are sufficiently accurate and the integration does not depend on the value of $R_{\epsilon}$. The expansions of the metric functions, with the nontrivial coefficients expressed in terms of the functions in $\rb = 0$ are
\begin{align}
     &\alpha = \alpha_0 + \frac{4 \pi}{3}\left(\rho_{0}+3 p_{0}\right) \rb^{2}+\mathcal{O}\left(\rb^{3}\right),\\ 
     &M = \frac{4 \pi}{3} \rho_{0} \rb^{3}+\frac{4 \pi}{5} \rho_{2} \rb^{5}+\mathcal{O}\left(\rb^{6}\right),\\
     &p = p_{0}-\frac{2 \pi}{3}\left(\rho_{0}+p_{0}\right)\left(\rho_{0}+3 p_{0}\right) \rb^{2}+\mathcal{O}\left(\rb^{3}\right),\\
     &\rho = \rho_{0}+\rho_{2} \rb^{2} + \mathcal{O}\left(\rb^{3}\right),\\
     &\chi = 1 - \frac{1}{2}\chi^{(2)} \rb^2 +\mathcal{O}\left(\rb^{3}\right) \text{(Exact case)},\\ 
     &\omega = \omega_{0}+\frac{8 \pi}{5}\left(\rho_{0}+p_{0}\right) \omega_{0} \rb^{2}+\mathcal{O}\left(\rb^{3}\right), \\ 
     &h_2 = h_2^{(2)} \rb^{2} + \mathcal{O}\left(\rb^{3}\right), \\
     &m_0 = \frac{8e^{-\alpha_0}}{5}\omega_0^2\left[ \frac{2\pi}{9}\left( 2\rho_0 + 3p_0 \right) - \frac{3}{8}\frac{\rho_2}{\rho_0 + 3p_0} \right]\rb^5 + \mathcal{O}\left(\rb^{6}\right), \\
     &\zeta_0 = \frac{3\omega_0^2e^{-\alpha_0}}{8\pi\left( \rho_0 + 3p_0 \right)}\rb + \mathcal{O}\left(\rb^{2}\right),
    \label{InitialConditions}
\end{align}

where $\omega_0$, as $p_0$, is an input parameter when it comes to solve the system. This parameter will determine the angular velocity of the star $\Omega_*$, as can be seen from the matching condition of $\omega$ with the exterior solution, $\omega^{\text{ext}}$, obtained in the previous section. This matching condition is simply given by imposing that the metric function $\omega$ and its first derivative are continuous throughout the star surface \cite{Yagi_2013}, i.e.
\begin{equation}
    \omega(R_*)=\omega^{\text{ext}}(R_*), \qquad \omega'(R_*)=\omega'^{\,\text{ext}}(R_*).
    \label{omegamatching}
\end{equation}
Therefore, in the rotating case, the stars are identified by a two-parameter family ($\omega_0$, $p_0$). The values of $\rho_2$ and $N_2$ are easily obtained from the EoS ($\rho(p)$, $n(p)$), and $h^{(2)}$ is obtained from \eqref{hEq}. The functions $k_2$ and $m_2$ are found to satisfy $k_2^{(2)} = m_2^{(2)} = -h_2^{(2)}$, around the center. As we have said, the odd powers in $\rb$ of almost all the metric functions are null, however the definitions of $M$ \eqref{Mdef}, $m$ and $\zeta_0$ \eqref{nicemetric} lead to the expansions given in \eqref{InitialConditions}.

Now we start the integration with a non-zero, but still unknown, seed for the second-order functions $h_2$ and $k_2$. To solve the unknown initial condition issue we will follow the approach given in \cite{Yagi:2013mbt, HartleI}. First we must obtain a particular solution for $h_2$ and $k_2$ ($h_p$, $k_p$) by solving the equations \eqref{hdot} and \eqref{kdot} for an arbitrary initial value (that must satisfy the regularity conditions given in \eqref{InitialConditions}). Next, we obtain a homogeneus solution ($h_h$, $k_h$) again for an arbitrary initial condition, using the same equations but with vanishing source terms. With these two functions we can construct the solution
\begin{equation}
          h_2(\rb) = h_p(\rb) + B h_h(\rb),\quad 
          k_2(\rb) = k_p(\rb) + B k_h(\rb).
\end{equation}

In these expressions $B$ is a constant that can be obtained by matching the functions $h_2$ and $k_2$ at the surface of the star with their corresponding exterior solutions. This matching condition is simply given by continuity of both functions at $R_*$, i.e.
\begin{equation}
    h_2(R_*)=h_2^{\text{ext}}(R_*),\quad  k_2(R_*)=k_2^{\text{ext}}(R_*).
\end{equation}
By doing this we are introducing the integration constant that appears in \eqref{ExteriorSol}, hence we have an algebraic system of two equations that can be solved for $A$ and $B$.

On the other hand, to solve the $l=0$ sector of the second order system, we first solve the coupled ODEs for $\zeta_0$ and $m_0$ as explained in the previous section, and then we obtain the solution for $h_0$
 up to a constant $h_0^{c}$ whose value is determined from the matching conditions 
\begin{subequations}
\begin{align}
&m_0^{\rm int}(R_*)-4\pi R_*^2\rho(R_*)\zeta_0(R_*)=m_0^{\rm ext}(R_*),\label{m0matching}\\
    &h_0^{\rm int}(R_*)=h_0^{\rm ext}(R_*), \label{h0matching}
\end{align}
\end{subequations}
 where the constant term in \eqref{m0matching} is due to a nonvanishing energy density at the surface of the star, as pointed out in \cite{Reina_2015,ReinaCompletion}. To obtain this constant term we can integrate \eqref{MasterEq} in the interval $\left[R_*-\epsilon, R_*+\epsilon \right]$ and take $\epsilon \rightarrow 0$. By doing this we have that all terms in the surface of the star vanish but the term $d\rho/dr$, which is unbounded at $R_*$ and contributes with a constant term. From \eqref{m0matching} we can obtain the value of $\delta M$, which reads
 \begin{equation}
   \delta M=m^{\rm int}(R_*)+\frac{J^2}{R^3_*}-4\pi R_*^2\rho(R_*)\zeta_0(R_*). 
   \label{deltaM}
 \end{equation}

We would also like to remark a subtle detail concerning the second-order equations. When solving the TOV numerically, the metric function $\alpha$ is not fixed to its correct initial value since it does not affect the observables of the star. However, for solving the second order problem it is necessary to find the correct initial value of this function since the second order perturbation functions depend directly on $\alpha(0)$ and an incorrect value will affect the values of the quadrupole moment and gravitational mass correction of the star. This can be done by first solving the TOV system, finding the correct initial value of $\alpha$ using the matching condition \eqref{Matchingzerothorder} and then solving both zeroth and second order systems.

\subsection{Global properties of compact stars}
A key feature of the Hartle-Thorne perturbative formalism is that it allows us to obtain the values of these observable parameters from the solutions of the Einstein equations for the interior of the star at each order in the expansion parameter. Indeed, once these solutions have been obtained, they can be matched to the exterior solutions, from which observational parameters such as the quadrupole moment can be obtained systematically.

Take for example the moment of inertia  $I$, which is defined as the quantity measuring how fast a star can spin
given a fixed spin angular momentum $J$, and is given by
\begin{equation}
    I=\frac{J}{\Omega_*}.
\end{equation}
To obtain the value of $I$ for a given (interior) solution of the second-order Hartle-Thorne equations is straightforward:  we simply obtain $J$ from \eqref{omegaext}, by matching the exterior solution to the interior one at $R_*$ \eqref{omegamatching} and dividing by $\Omega_*$. It will be convenient also to define the dimensionless moment of inertia as
\begin{equation}
    \bar I=\frac{I}{M^3_*}.
\end{equation}

On the other hand, the metric generated by an isolated, static gravitating body at a given point sufficiently far from the source may be written using a multipolar expansion in a system of \emph{Asymptotically Cartesian and Mass Centered} coordinates \cite{KThorneMultipolar,HartleThorne85}, whose (0,0) component will be of the form
\begin{equation}
g_{00}= -1 + \frac{2M}{r}+3\frac{Q_{ij}}{r^5}x^ix^j+\order{\frac{1}{r^4}},
\label{ACMC expansion}
\end{equation}
 where $M$ is the gravitational mass of the star \footnote{In stationary spacetimes, the gravitational mass is defined via a Komar surface integral \cite{Gourgoulhon:2010ju}, and coincides with the ADM mass in asymptotically flat spacetimes.},
 and $Q_{ij}$ is the (traceless) quadrupolar tensor. 

The induced quadrupolar deformation of the star can be described in terms of the star’s $l=2$ sector perturbation functions in spherical coordinates. Indeed, defining $x^i=rn^i(\theta,\phi)$, where $n^i$ is the unit three-vector in spherical coordinates, we may write:
\begin{equation}
     Q_{ij}\frac{x^ix^j}{r^2}=Q_{ij}n^in^j(\theta,\phi)=\sum_{m=-2}^2 Q^m Y_{2m}(\theta,\phi),
\end{equation}
(where $Y_{2m}$ are the $l=2$ spherical harmonics). We find, in the case of an axially symmetric deformation, that the expansion \eqref{ACMC expansion} reduces to
\begin{equation}
    g_{00}= -1 + \frac{2M}{r}+\frac{Q}{r^3}P_2(\cos\theta)+\order{\frac{1}{r^4}},
    \label{Quadrupolemoment}
\end{equation}
which defines the quadrupole moment of the metric, $Q$.

Thus, we may perform an asymptotic expansion of the Hartle-Thorne perturbative solution for the exterior spacetime metric and identify the gravitational mass and quadrupole moment as the coefficients proportional to $2/r$ and the $P_2(\cos\theta)/r^3$ term, respectively.
Clearly, these quantities get corrections due to the star rotation.

Indeed, for example, the gravitational mass of the star, up to second order in $\eps$, receives a correction
\begin{equation}
    M(\eps)= M_*+\eps^2\delta M
\end{equation}
which can be obtained from the expansion of the $h^{\rm ext}_0$ perturbation function. Furthermore, taking into account the asymptotic expansion for large $r$ of $h_2^{\text{ext}}$ and $\omega^{\text{ext}}$ one finds that the  spin-induced quadrupole moment of the star, up to second order in the spin parameter, is given by
\begin{equation}
    Q^{\text{rot}}=-\frac{J^2}{M_*}- \eps^2 \frac{8}{5}AM^3_*.
\end{equation}
For later convenience we also define the dimensionless rotationally-induced quadrupole moment as
\begin{equation}
    \bar{Q}^{\text{rot}}=-\frac{M_*}{J^2}Q^{\text{rot}}.
\end{equation}
Dropping the staticity assumption, nontrivial current multipole moments may appear in the expansion of the $(0,j)$ components of the metric,
\begin{equation}
    g_{0j}=-2\epsilon_{jkl}\frac{J_k}{r^2}x^l-4\epsilon_{jqk}\frac{S^k_l}{r^5}x^qx^l+\order{\frac{1}{r^4}},
    \label{multipolecurrents}
\end{equation}
the first term corresponding to a non-vanishing angular momentum.

Finally, another interesting property that can be obtained from the solutions is the binding energy, which physically corresponds to the amount of energy that keeps all the particles (baryons) in the star from dispersing to infinity. It is defined as
$
    E_t=M_g-M_b
$
where $M$ is the gravitational mass (in the static case, $M=M_*$) and $M_b$ is the baryon mass of the star. The binding energy so defined includes both the gravitational binding energy and the nuclear binding energy.  However, we will be mostly interested in the gravitational contribution to the total binding energy, i.e. the \emph{gravitational} binding energy, since it contains EoS-independent information about the mass distribution of the star \cite{UniversalBinding}. The gravitational binding energy is defined as $E_g= M-M_p$, being $M_p$ the proper mass, given by the proper energy-momentum density, $P_\mu=T_{\mu\nu}u^\nu$, integrated on a spacelike hypersurface with volume form $dS_\mu$:
\begin{equation}
M_p=\int_\Sigma T^\mu_{\nu}u^\nu dS_\mu.
\end{equation}
In a stationary spacetime, this integral does not depend on the chosen hypersurface, so we may take $dS_\mu=n_\mu d^3S$, where $d^3S=\sqrt{\gamma}d^3x$ is the volume element of the spacelike hypersurfaces defined by $t=\text{const}$, $\gamma$ is the determinant of the three-metric associated with these hypersurfaces and $n_\mu=\nabla_\mu t/\sqrt{(\nabla_\nu t\nabla^\nu t)}$ is the corresponding normal vector, so that, for the static case,
\begin{equation}
	M_p^*=\int T^\mu_{\nu}u^\nu n_\mu d^3S=4\pi\int_0^{R_*}\!\!\frac{\rho(r) }{\sqrt{1-\frac{2M(r)}{r}}}r^2dr .
\end{equation}  

In the slowly rotating case, the perturbed proper mass $M_p(\eps)$ will also get corrections. Expanding both $\gamma$ and the product $u^\mu n_\mu$ in powers of $\eps$, we have, up to second order,  ${M_p(\eps)=M^*_p+\eps^2\delta M_p}$, where

\begin{equation}
    \delta M_p=8\pi \int_0^{R_*}\rho \rb^4 e^{\beta/2}\qty{\qty(\frac{m_0}{\rb-2M})+\frac{\rb^2}{6}\omega^2 e^{-\alpha/2}} d\rb.
    \label{delMp}
\end{equation}
Hence, it is straightforward to obtain the second order perturbation to the gravitational binding energy ${
    E_g(\eps)=E_g+\eps^2\delta E_g }
$, with $\delta E_g =\delta M-\delta M_p$.

%%%%%%%%%%%%%%%%%
\section{Tidally deformed stars and Love numbers. }
%%%%%%%%%%%%%%%%%%%
\label{Sec:Tid.demform}
Until now we have studied the deformation of stars resulting from their own rotation. However, we can also study (non-rotating) stars which are deformed due to some external tidal force. Tidal forces are one of the principal signatures of the presence of a nontrivial gravitational field in spacetime. Such forces are responsible for relative acceleration among freely falling particles. This acceleration induces, on extended gravitating bodies, a field of strains that causes a deformation, which may be measured. By measuring the deformation response of a body to a tidal gravitational field, we may obtain information about the kind of matter that conforms the body, as well as its equation of state. In particular, in the case of binary systems involving neutron stars, it is very useful to analyze the deformation of the stars due to tidal effects, which may be measured from its gravitational wave spectrum previous to the merging.

On the other hand, as we have previously stated, a spherical body immersed in an external tidal field may deform due to tidal forces. Owing to this deformation, the metric in the exterior spacetime will develop a non trivial multipolar structure.
To characterize the tidal field generated by a given source, consider an observer immersed in a tidal field generated by an external source. We may expand the metric of spacetime in a region surrounding the observer's worldline in Fermi normal coordinates, with the $(0,0)$ and $(0,j)$ component of the metric given by \cite{HartleThorne85, Poisson_Motionpointparticles},
\begin{subequations}
\label{Fermicoordexpansion}
\begin{align}
        g_{00}=-1+\mathcal{E}_{ij}x^ix^j+\order{r^3},\label{Fermicoordexpansion1}\\
        g_{0j}=\frac{2}{3}\epsilon_{jqk}\mathcal{B}^q_lx^kx^l+\order{r^3}\label{Fermicoordexpansion2},
\end{align}
\end{subequations}
where $\mathcal{E}_{ij}$ and $\mathcal{B}_{ij}$ are the (quadrupolar) tidal multipole moments of electric and magnetic type, respectively. These two are related to the Riemann tensor through $\mathcal{E}_{ij}=R_{i0j0}$ and $\mathcal{B}^i_j=\frac{1}{2}\epsilon^{ijk}R_{0jkl}$ \cite{HartleThorne85}. The quadrupolar tidal moments are independent of the distance to the source, but may depend on the time coordinate if the source is not stationary.
Now, instead of the worldline of an observer, we may consider the worldtube of an extended, gravitating body immersed in an external tidal field. We thus may be able to write the (0,0) component of the metric outside this body by combining both \cref{ACMC expansion,Fermicoordexpansion1},
\begin{equation}
     g_{00}=-1+ \frac{2M}{r}+3\frac{Q_{ij}}{r^5}x^ix^j+\order{\frac{1}{r^4}}+\mathcal{E}_{ij}x^ix^j+\order{r^3},
     \label{general expansion tt}
\end{equation}
whereas the $(0,j)$ component of the metric will be given by the combination of \cref{multipolecurrents,Fermicoordexpansion2},
\begin{equation}
     g_{0j}= -4\epsilon_{jqk}\frac{S^k_l}{r^5}x^qx^l+\order{\frac{1}{r^4}}+\frac{2}{3}\epsilon_{jqk}\mathcal{B}^q_lx^kx^l+\order{r^3}.
     \label{general expansion tj}
\end{equation}

Note that, by writing the metric as in \cref{general expansion tt,general expansion tj}, we are assuming that there exists a region of the exterior spacetime, called the \emph{buffer region}, in which the expansions of \cref{ACMC expansion,multipolecurrents,Fermicoordexpansion} converge simultaneously. This will be well justified in the limit in which the source of the external tidal field is very far away from the body that gets deformed and does not evolve rapidly with time. It can also be shown that, in this limit, the multipole moments appearing in \eqref{general expansion tt} are defined unambiguously \cite{Equivalencemultipoles}.

%\subsection{Tidal deformability and Love numbers}

\subsection{Electric quadrupolar Love number}
Since we are considering that the body gets deformed due to the external tidal field, the quadrupole tensor $Q_{ij}$ will be a more or less complicated function of the tidal field $\mathcal{E}_{ij}$. However, working to linear order in the tidal moment, we define the (tidal) electric quadrupolar deformability $\lambda_t$ as
\begin{equation}
Q_{ij}=-\lambda_t \mathcal{E}_{ij}.
\label{TidalDefinition}
\end{equation}
 Assuming that the terms with non-zero axial number $m$ vanish, we may write \eqref{Fermicoordexpansion1} in spherical coordinates as
\begin{equation}
       g_{00}=-1+r^2\mathcal{E}P_2(\cos\theta)+\order{r^3},
       \label{Expns}
\end{equation}
so that the tidal electric Love number can be obtained as the ratio $\lambda_t=-Q/\mathcal{E}$, where $Q$ is the quadrupole moment of the star as defined in \eqref{Quadrupolemoment}.

That the deformation of the star resulting from an external tidal field will be well described by its deformability $\lambda$ is consistent with the assumption that the source of this external field is far from the body, since the tidal field will be weak and the linear approximation will be well justified.

The quadrupolar deformation of the star due to an external tidal field and to a slow rotation  can be described by a similar spacetime metric (up to second order) \cite{Hinderer:2007mb,Hinderer:2009ca}, hence we can take advantage of the differential equations derived above to obtain the results for a tidally deformed star.
Indeed, to describe a tidally deformed star, one introduces the  metric perturbation $h_{\mu\nu}$ as in \cref{Appendperturb}. By direct comparison between the metric \eqref{perturbmetric} and \eqref{MetricperturbationPleg}, it is straightforward to see that the $l=2$ even perturbation functions $H_2,M_2$ and $K_2$ in the tidally deformed case play a similar role as the functions $h_2,m_2,k_2$ in the slowly rotating case.  Indeed, this can be seen by redefining these functions as 
\begin{equation}
    H_2=2e^\alpha h_2,\quad M_2=2e^{\beta} \frac{ m_2}{\rb-2M}.
\end{equation}
In order to calculate the quadrupolar deformation of the metric due to an external gravitational field, the odd perturbations to the metric are not needed. Therefore, the metric functions of non-rotating tidally deformed stars can be directly obtained from \cref{mdos,kdot,hdot} by imposing $\omega = 0$. Then, these equations can be arranged into only one equation for $h_2$,
\begin{equation}
    \begin{aligned}
         h''_{2} = & -\left\{\frac{2}{\rb} + \left[\frac{2 M}{\rb^2} + 4 \pi \rb(p-\rho)\right] e^{\beta}\right\} h'_{2}\\
        &+\left\{\frac{6 e^{\beta}}{\rb^{2}}-4 \pi\left[5 \rho+9 p+(\rho+p) \frac{d \rho}{d p}\right] e^{\beta}\right.  \left.+\left(\alpha'\right)^{2}\right\} h_{2}.
        \label{MasterEq}
    \end{aligned}
\end{equation}
This is a second-order differential equation which can be solved as a first-order system as in \cite{Hinderer:2007mb} by defining $H\equiv h'_2$,
while $H'$ is given by \eqref{MasterEq}. To do this we need, again, to expand the function $h_2$ in powers of $\rb$ and introduce it in \eqref{MasterEq} to obtain the initial condition
\begin{equation}
    \begin{aligned}
        &h_2 = h_2^{(2)}\rb^2  + \mathcal{O}\left(\rb^{3}\right)\\
        &H = h'_2 = 2h_2^{(2)}\rb  + \mathcal{O}\left(\rb^{2}\right).
        \label{InitTidal}
    \end{aligned}
\end{equation}
Once more, the value of $h^{(2)}$ can not be found from the limit $\bar r\rightarrow 0$ of the field equations. However, we will see that this is unimportant to find the correct value of the tidal deformability (see below), so that we can start the integration with an arbitrary value for $h_2^{(2)}$, as long as  \eqref{InitTidal} are satisfied.

Once the internal solution has been found numerically, we can calculate the (tidal) Love number from the external solution of the metric after matching with the internal solution using the matching conditions
\begin{subequations}
\begin{align}
    h_2^{\rm int}(R_*)=h_2^{\rm ext}(R_*),\\ H^{\rm int}(R_*)-4\pi\frac{R_*^2}{M_*}\rho(R_*)h^{\rm int}_2(R_*)=H^{\rm ext}(R_*).\label{Hmatching}
\end{align}
\end{subequations}
where, as in the case of \eqref{m0matching}, there is a constant contribution in \eqref{Hmatching} due to a nonvanishing energy density at the surface of the star \cite{Hinderer:2009ca,Yagi:2013mbt,Damour:2009vw}.

Hence, as before, the vacuum ($\rho = p = 0$) version of \eqref{MasterEq} can be analytically solved,
\begin{widetext}
\begin{equation}
    \begin{aligned}
        h_{2}^{\text {ext }}=\: c_{1}\left(\frac{r}{M_{*}}\right)^{2}&\left(1-\frac{2 M_{*}}{r}\right)
         \times\left[-\frac{2 M_{*}\left(r-M_{*}\right)\left(3 r^{2}-6 M_{*} r-2 M_{*}^{2}\right)}{r^{2}\left(r-2 M_{*}\right)^{2}}+3 \ln \left(\frac{r}{r-2 M_{*}}\right)\right]+c_{2}\left(\frac{r}{M_{*}}\right)^{2}\left(1-\frac{2 M_{*}}{r}\right).
    \end{aligned}
\end{equation}
\end{widetext}
where $c_{1,2}$ are constants that can be determined through the matching conditions in terms of $H^{\rm int}(R_*),h^{\rm int}_2(R_*)$.
Studying the behavior of this function in the buffer zone we can extract the expression for $Q$ and $\mathcal{E}$ in terms of these constants $c_1$ and $c_2$ in order to obtain the tidal electric deformability. Indeed, in the buffer zone
\begin{equation}
    h_{2}^{\text {ext }}=\frac{16}{5} c_{1} \frac{M_{*}^{3}}{r^{3}}+c_{2} \frac{r^{2}}{M_{*}^{2}}+\mathcal{O}\left(\frac{M_{*}^{4}}{r^{4}}, \frac{r}{M_{*}}\right),
\end{equation}
and comparing with \cref{Quadrupolemoment,Expns} we have 
\begin{equation}
    \lambda_t=\frac{16}{15}M^5\frac{c_1}{c_2},
\end{equation}
and, defining the tidally induced, quadrupolar electric Love number $k^{E}_2=\tfrac{3}{2} \tfrac{\lambda_t}{R_*^5}$, we can write
\begin{align}
        &k_{2}^{E} = \frac{8}{5} C^{5}(1-2 C)^{2}[2+2 C(y-1)-y]\times\nonumber\\[2mm]&
        \{2 C[6-3 y 
        +3 C(5 y-8)]+4 C^{3}\left[13-11 y+C(3 y-2)\right.\nonumber\\[2mm]&\left.\!\!+2 C^{2}(1\!+\!y)\right]
        \left.\!\!+3(1\!-\!2 C)^{2}[2\!-\!y+2 C(y\!-\!1)] \ln (1\!-2 C)\right\}^{-1}\!\!,
    \end{align}
where $y = R_* H^{\rm ext}(R_*)/h^{\rm ext}_2(R_*) $ and $C=M_*/R_*$ is the compactness of the zeroth-order solution. From the definition of $y$ it is clear that $k^{\text{(tid)}}_2$, and hence $\lambda_t$, does not depend on the value of $h^{(2)}$ chosen in the numerical integration of the interior equation. Again, for later convenience we will define the adimensional tidal deformability as
\begin{equation}
    \bar{\lambda}_t = \frac{2}{3}k_2^{\text {(tid)}}C^{-5}.
\end{equation}

%%%%%%%%%%%%%%%%%
\subsection{Magnetic quadrupolar Love number}
%%%%%%%%%%%%%%%%%
 While electric-type Love numbers measure the induction of different multipole moments on a star due to an external gravitational field and can also be calculated in the Newtonian limit of general relativity, the current multi-pole moments induced by an external magnetic-type tidal field have no analogue in Newtonian gravity, and thus the magnetic tidal Love numbers are a genuine prediction of general relativity. In the simplest (quadrupolar) case, the tidal magnetic deformability, in analogy with the electric case, measures the magnitude of the quadrupolar current $S_{ij}$ induced in the star by an external tidal field of magnetic type, $\mathcal{B}_{ij}$. At the linear level, the relation between both is 
 \begin{equation}
     S_{ij}=-\sigma_t\,\mathcal{B}_{ij}.
 \end{equation}
  Therefore, it is interesting to study the response of a NS under a magnetic-type external gravitational field, whose effects may be relevant for such compact objects. To do so, we consider an axially-symmetric perturbation of the spherical metric.  For the calculation of magnetic-type Love numbers, only the odd metric perturbations are relevant. The magnetic Love number can be obtained by assuming a perturbation of the static metric of the form
$
     g(\epsilon)=g^{(0)}+\epsilon h^{\rm odd}
$
where $g^{(0)}$ is the static spherically symmetric metric \eqref{staticmetric}, $\epsilon$ here does not have to do anything with rotation, but will play the role of a bookkeeping parameter, and $h^{\rm odd}$ is the odd-parity perturbation:
\begin{equation}
    h^{\rm odd}_{\mu\nu}dx^\mu dx^\nu=2V(r,\theta)drd\phi +2\omega(r,\theta)dtd\phi.
\end{equation}
In particular, for the simplest (quadrupolar) perturbations, we take into account only $l=2$ in \eqref{MetricperturbationPleg} and we have
\begin{equation}
\begin{split}
      V(r,\theta)=  V_2(r)\sin\theta\, \partial_\theta P_2(\cos\theta),\\ \omega(r,\theta)=\omega_2(r)\sin\theta\, \partial_\theta P_2(\cos\theta).
      \label{oddperturbPleg}
\end{split}
\end{equation}
Notice that we have dropped the barred radial coordinate, since the star shape will not be altered by the odd metric perturbations.

On the other hand, using the notation of \cref{Appendperturb}, we define $n^i_A=\partial_A n^i$. Then, we can transform the $(0,j)$ components of the metric to spherical coordinates by $g_{0j}\rightarrow g_{0A}=rn^j_A g_{0j}$, and expand into odd-parity vector harmonics (see \cref{Appendperturb}), for instance,
\begin{equation}
   \epsilon_{jqk}\frac{S^k_l}{r^3}x^qx^l\rightarrow n^j_A\epsilon_{jqk}S^k_ln^qn^l=\sum\limits_{m=-2}^{m=2}S_{m}X_A^{2m}(\theta,\phi),
\end{equation}
so that \cref{general expansion tj} is transformed into
\begin{equation}
     h^{\rm odd}_{0A}= \sum\limits_{m=-2}^{m=2}\qty{\frac{-4}{r^2}S_{m}+\frac{2}{3}r^3B_{m}}X_A^{2m}(\theta,\phi)+\cdots
     \label{general expansion vetor harm}
\end{equation}
In particular, for the simplest case of axially symmetric perturbations, we have, for instance,
\begin{equation}
     h^{\rm odd}_{0\phi}= \qty[\frac{-4S}{r^2} +\frac{2}{3}r^3B]\sin{\theta}\partial_\theta P_2(\cos\theta)+\cdots ,
     \label{general expansion Pleg}
\end{equation}
where $\cdots$ in \cref{general expansion vetor harm,general expansion Pleg} denotes the non-leading terms in the expansion at the buffer zone. 
Hence, the magnetic tidal deformability can be obtained as the ratio 
\begin{equation}
    \sigma_t=-\frac{S}{B},
\end{equation}
where the constants $S$ and $B$ will be determined from the buffer zone expansion of the odd-parity metric  perturbation functions \eqref{oddperturbPleg}. 

To find these functions, the Einstein equations must be solved for the star interior, and matched to a suitable exterior solution, as we have previously done for the even parity case. However, in the case of odd-parity tidal perturbations, the energy-momentum tensor of the fluid will also get perturbed through a perturbation of the 4-velocity, $u^\mu(\epsilon)=u^\mu+\epsilon\delta u^\mu$,
\begin{equation}
    \delta {T^\mu}_\nu=(\rho+p)(u_\nu\delta u^\mu +u^\mu\delta u_\nu)-p g^{(0)\mu \sigma}h^{\rm odd}_{\sigma\nu}
\end{equation}
working to first order in $\epsilon$, and $\delta u_\mu=g^{(0)}_{\mu\nu}\delta u^\nu+h^{\rm odd}_{\mu\nu}u^\nu$.

Now, in principle, the four-velocity perturbations are independent of the metric perturbations, and the latter are only related with the former  through the perturbative Einstein equations. However, there are two simple cases for which these perturbations are closely related to each other: the \emph{static} case, in which the fluid remains static --- with vanishing spatial four-velocity components  i.e. $\delta u^\mu=0$ --- even when the metric perturbations are taken into account, and the \emph{irrotational} fluid, which is based on the assumption that the fluid perturbations preserve the relativistic circulation theorem \cite{rezzolla2013relativistic}, and can be shown to be equivalent to the condition of a vorticity-free fluid, i.e. with vanishing vorticity four-vector $\omega^\alpha=\frac{1}{2}\epsilon^{\alpha\beta\mu\nu}u_\beta\nabla_\mu u_\nu=0$ which in turn implies $\delta u_\mu=0$ \cite{GravitomagneticIrrot}, since the static initial configuration is trivially vorticity-free. The latter assumption is usually considered as more physically relevant, as the static fluid is only adequate for the non-physical case of time-independent tidal perturbations \cite{MagneticLoveClarified}. Hence in the following we will only consider the case of an irrotational fluid and write the four-velocity perturbation as
\begin{equation}
    \delta u^\mu = {g^{(0)}}^{\mu\nu}h_{\nu\alpha}^{\rm odd}u^\alpha = -\frac{2\omega_2(r)}{r^2}\frac{\partial_\theta P_2(\cos\theta)}{\sin\theta}\delta_\phi^\mu.
    \label{4veloddperturbations}
\end{equation}
Substituting this expression into the stress-energy tensor perturbation, we may expand the Einstein equations as in \eqref{PertEinstein} and solve the linearized equations $E^{(1)}=0$, which yield $V_2=0$ and the following ODE:
\begin{align}
{\omega}_{2}^{\prime \prime}&\!-4 \pi(\rho\!+\!p) re^{\beta}{\omega}_{2}^{\prime}-\!\frac{e^{\beta}}{r^{3}}\left(6r\!-\!4 M\!-\!8 \pi(\rho\!+\!p) r^{3}\right) \!{\omega}_{2}\!=\!0
\label{Oddome2}
\end{align}
for the metric perturbation function $\omega_2$. The above equation is numerically integrated in the star interior starting with an initial condition
\begin{equation}
    \omega_2= \omega_2^{(3)}r^3+\order{\bar{r}^5}
\end{equation}
where, as in the electric case, the exact value of $\omega_2^{(3)}$ is undetermined from the equations but won't be needed for obtaining the corresponding Love number. 

On the other hand, the exterior solution of \cref{Oddome2} can be written in terms of the hypergeometric function $_2F_1(\alpha,\beta,\gamma;x)$ as \cite{MultipoleLove}
\begin{equation}
\begin{split}
    \omega_2^{\rm ext}(r)=&d_1\qty(\frac{r}{2M})^3\,_2F_1\qty(-1,-4,-4;\frac{2M}{r})+\\+&d_2\qty(\frac{2M}{r})^2\,_2F_1\qty(1,4,6;\frac{2M}{r})
    \label{ome2ext}
\end{split}
\end{equation}
where $d_{1,2}$ are integration constants that can be related to the interior solution through the matching conditions 
\begin{equation}
    \omega_2^{\rm ext}(R_*)=\omega_2(R_*),\quad {\omega'}_2^{\rm ext}(R_*)=\omega'_2(R_*).
\end{equation}
The expansion of \eqref{ome2ext} in powers of $r$ and $r^{-1}$ in the buffer region reads
\begin{equation}
      \omega_2^{\rm ext}(r)= \frac{1}{8}d_1\frac{r^3}{M_*^3}+4d_2\frac{M_*^2}{r^2}+\order{\frac{M_{*}^{3}}{r^{3}},\frac{r^2}{M^2_{*}}},
\end{equation}
And, comparing with \eqref{general expansion Pleg}, we have

\begin{equation}
    \sigma_t=\frac{16}{3}\frac{d_1}{d_2}M^5
\end{equation}
so that the magnetic quadrupolar Love number ${k_2^{M}=48\tfrac{\sigma_t}{R_*^5}}$ reads
\begin{equation}
%\hspace*{0.7cm}
\begin{split}
k_{2}^{M}&=\frac{96 C^{5}}{5}[3+2 C(y-2)-y]\big\{2 C\{9-3 y+C[3(y-1)+\\+&2 C(C+\!y\!+C y)]\}
\!+\!3[3+2 C(y-2)-y] \log (1-2 C)\big\}^{-1}\!\!.
\end{split}
\end{equation}
where now $y = R_* \omega'_2(R_*)/\omega_2(R_*) $, and once more it is clear from this expression that the initial condition for $\omega_2$ does not enter in the expression for the magnetic Love number.
\section{Quasi-universal relations}
\label{sect:Quasi-Univ-rel}
\subsection{I-Love-Q}

In their original paper \cite{Yagi_2013}, Yagi and Yunes present a set of EoS-independent relations between the dimensionless moment of inertia, quadrupole moment and Love numbers of slowly rotating and tidally deformed compact stars, the so-called \emph{I-Love-Q} relations. Soon after these relations where proposed, in \cite{Yagi:2014qua} two possible reasons for these relations to exist were given. The first one relies on the fact that these relations depend mostly on the outer core ($10^{13} \leq \rho \leq 5\:10^{14} \: \text{g/cm}^3$) of the NS, where all the EoS extracted from the experimental data of the nuclear physics (SLy, APR, WFF1, etc.) follow the same behavior. The second is related to the no-hair conjecture of black holes (BH) since the three parameters $(\bar I,\bar\lambda$ and ${\bar Q)}$ must  approach the limiting  values of a BH for stars with large compactness. 

\begin{widetext}

\begin{figure}[h!]
%\hspace*{-1.0cm}
    \includegraphics[scale=0.7]{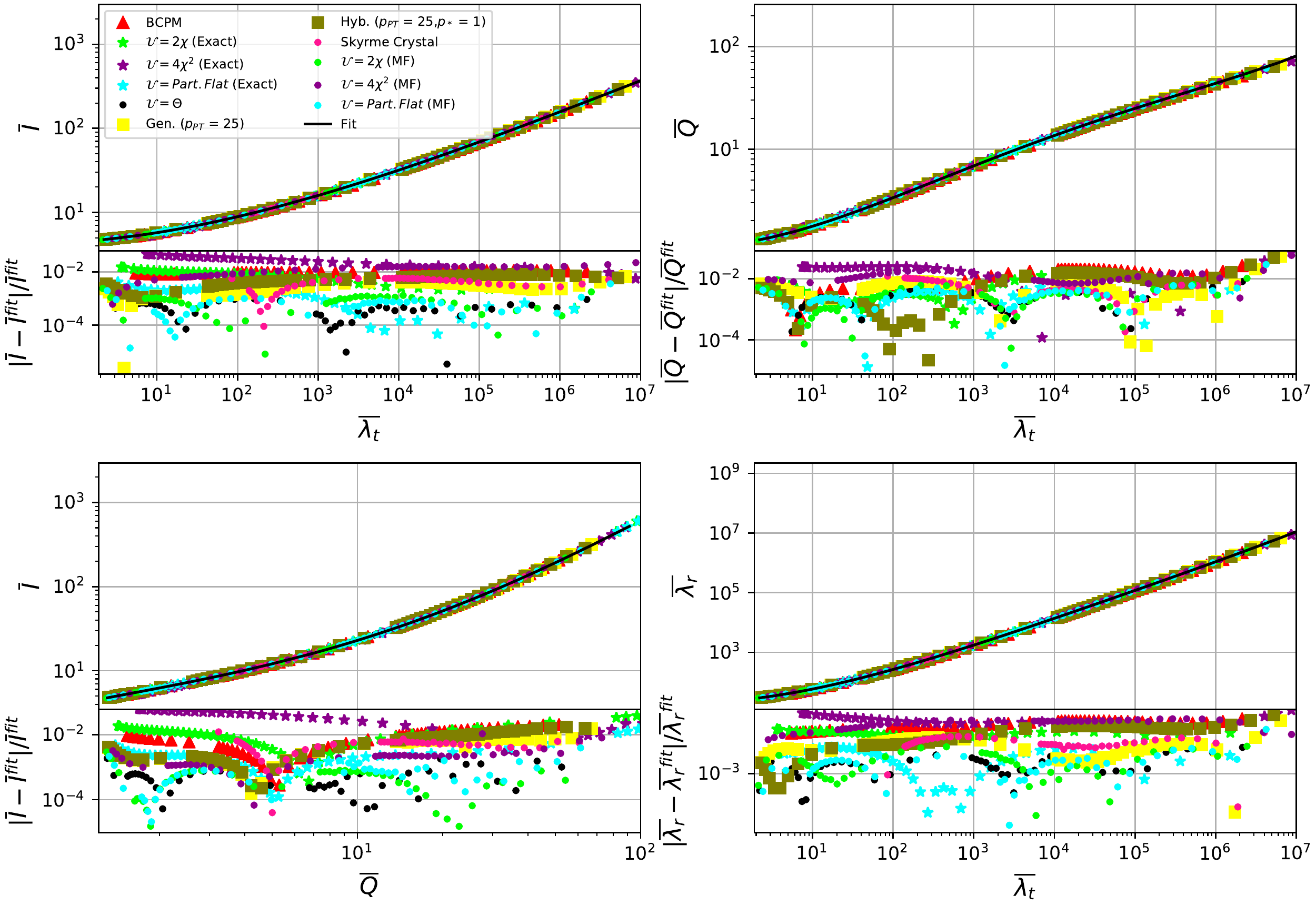}
    \centering
    \caption{I-Love-Q relations and relation between the rotationally induced and tidally induced deformabilities for different Skyrme models, both the exact and mean field solutions. The black line corresponds to the numerical fit obtained in \cite{Yagi_2013}.}
    \label{fig:ILoveQ}
\end{figure}

\end{widetext}

In \cref{fig:ILoveQ} we show that these relations for the BPS (both exact and mean-field limits) also satisfied. We also show the data for the standard Skyrme crystal, generalized and hybrid EoS of \cite{Adam:2020yfv} (which satisfy these relations as well) and the numerical fit for each of these relations obtained in \cite{Yagi_2013} is plotted with a black line. Although somewhat expected, this result is remarkable at least for the case of exact BPS models, for which the I-Love-Q relations are satisfied even when they present a non-barotropic EoS which varies depending on the chosen potential. Furthermore, the relations are satisfied for these models in the exact and mean-field cases. As we will see, this will not be true anymore for other quasi-universal relations. This points out the universality and EoS independence of the \emph{I-Love-Q} relations.

\begin{figure}
    \hspace*{-0.7cm}
    \includegraphics[scale=0.42]{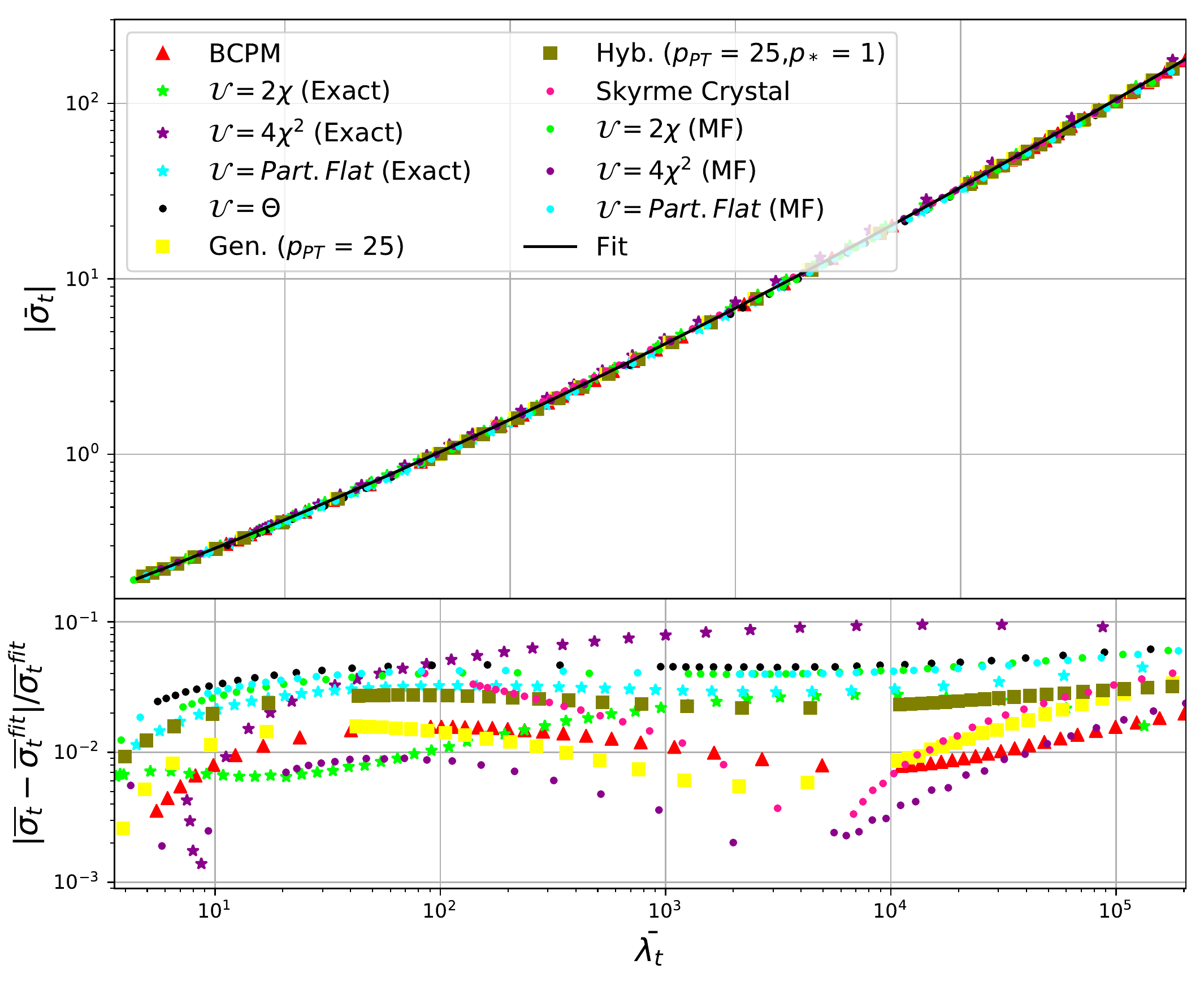}
    \caption{Quasi-universal relation between electric and magnetic quadrupolar deformabilities.}
    \label{fig:elec-magLove}
\end{figure}

Additionally, more quasi-universal relations have been found between electric, magnetic and higher multipole Love numbers \cite{MultipoleLove}. For example, in \ref{fig:elec-magLove} we show how there is as well an EoS independent relation between the (dimensionless) electric and magnetic quadrupolar tidal deformabilities in all models considered.

\subsection{I-Love-C}

\begin{figure}
    \hspace*{-0.7cm}
    \includegraphics[scale=0.4]{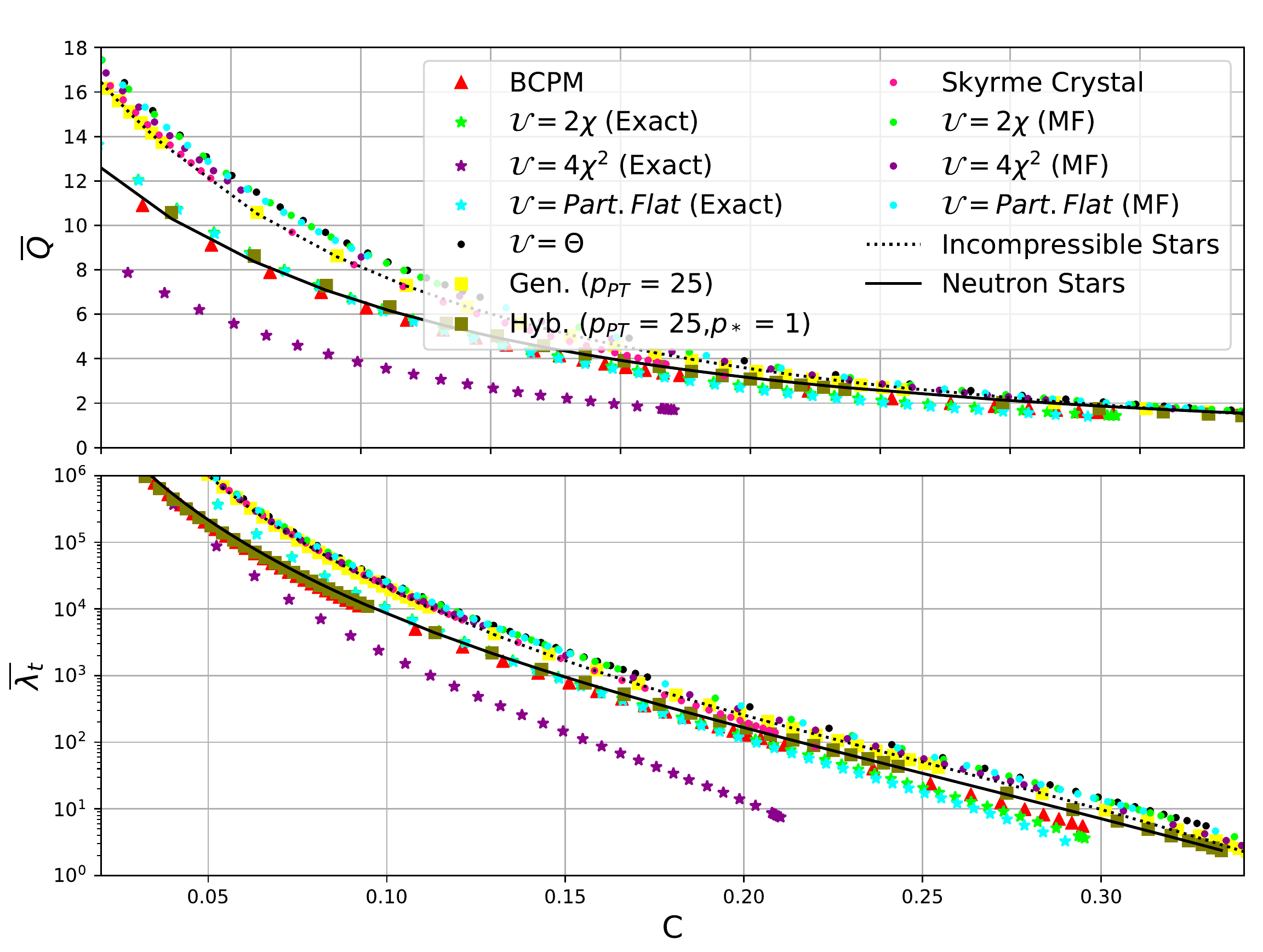}
    \caption{Relations of the dimensionless quadrupolar moment and electric quadrupolar Love number with compactness.}
    \label{fig:ILoveC1}
\end{figure}

\begin{figure}
    %\hspace*{-0.5cm}
    \includegraphics[scale=0.4]{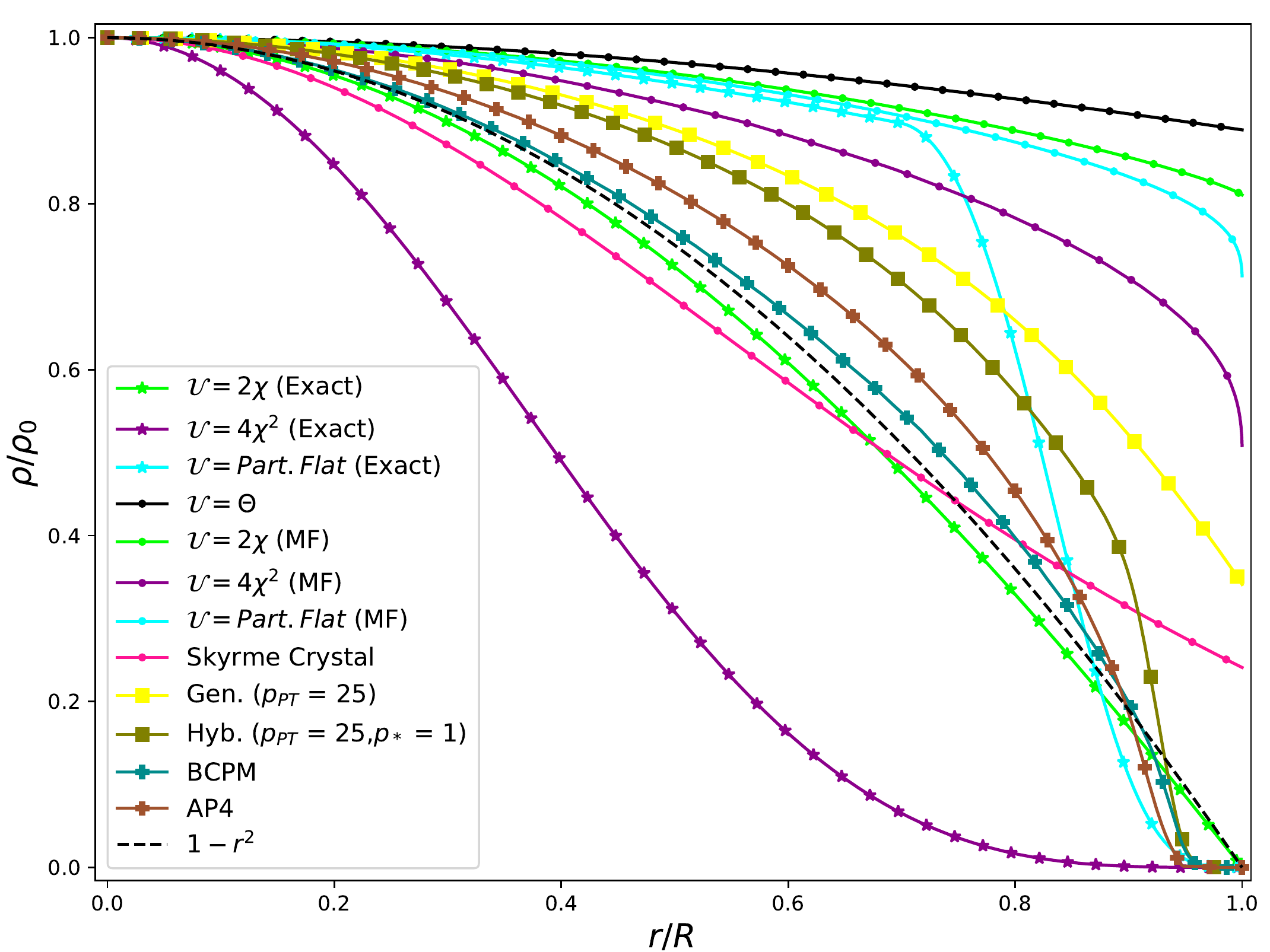}
    \caption{Normalized energy density profiles of $1.4 M_{\odot}$ Skyrme stars for different models. We include the nuclear based EoS BCPM and AP4, as well as the quadratic curve $1-r^2$. }
    \label{fig:RhoVSr}
\end{figure}

Apart from the \emph{I-Love-Q} relations, there exists other set of relations between the moment of inertia, the Love numbers and the compactness of neutron stars that share some chatracteristics with the \emph{I-Love-Q} but are accurate only up to $\sim 10\%$. These \emph{I-Love-C} relations were approximately derived analytically in \cite{I-Love-C}, as well as a possible explanation for these relations, in terms of the behavior of the energy density in the star interior.
It turns out that these relations, as opposed to the \emph{I-Love-Q} relations, are not universally satisfied for all the models we have considered. Indeed, from \cref{fig:ILoveC1,fig:ILoveC1} it can be seen that the relation between $I,Q$ and $C$ generally splits into two branches, corresponding to usual neutron stars and incompressible stars. This is consistent with the findings of \cite{I-Love-C}. However, we also find that, although the mean field version of the BPS models does lie in the incompressible star branch, the exactly solved cases behave quite differently. Whereas the behavior of the Partially-flat and $2\chi$-BPS models is better adjusted  by the NS branch, the $4\chi^2-$BPS model does not fit in neither branch. This behavior can be traced back to the radial dependence of the energy density in each model. 
\begin{figure}
    \hspace*{-0.3cm}
    \includegraphics[scale=0.4]{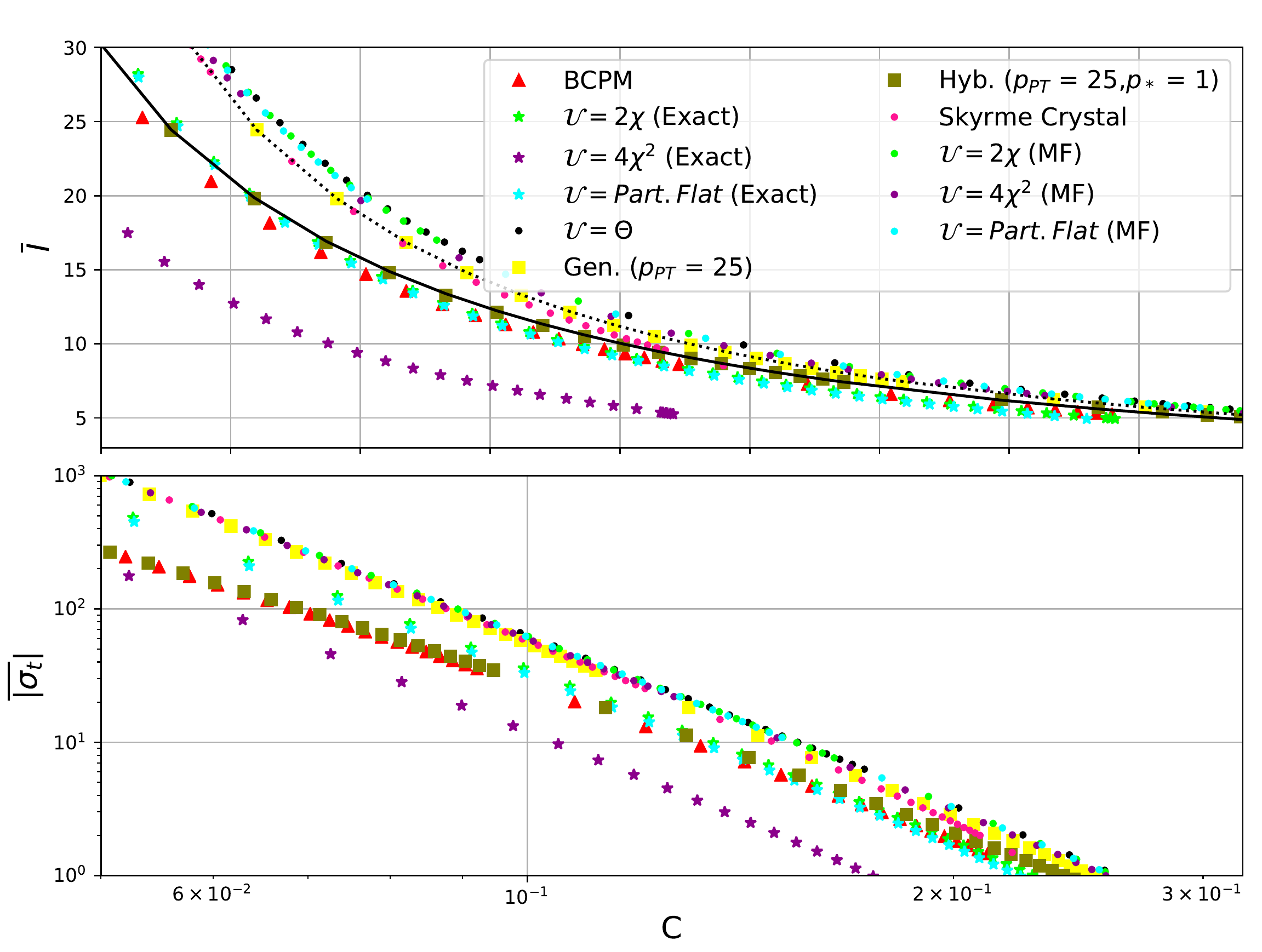}
    \caption{Relations of the dimensionless moment of inertia and magnetic quadrupolar Love number with compactness.}
    \label{fig:ILoveC2}
\end{figure}
Indeed, from \cref{fig:RhoVSr} we can see that the mean field approximation is not good in order to describe the low-density regime of neutron stars within the BPS Skyrme models in general, which translates into very different behaviors of the \emph{I-Love-C} relations for these models. Indeed, it is clear from this figure that the MF approach overestimates the energy density of the stars in the outer regions. 

Furthermore, the energy density profile for the different BPS models highly depends on the chosen potential. For example, while the $\theta$-potential yields almost incompressible stars, the  $2\chi$-potential curve can be well approximated by a quadratic function. This quadratic behavior is in fact expected for realistic neutron stars, whilst the behavior of the density profile for the $4\chi^2$-model is actually more similar to that of white dwarfs \cite{I-Love-C}. Indeed, as we have seen in \cref{fig:ILoveC1,fig:ILoveC2}, the (exact) $2\chi$-BPS model \emph{I-Love-C} relations are very close to the NS fit from \cite{I-Love-C}. Finally, it is interesting that the curve for the P.F. potential does not fit to a quadratic curve, yet the \emph{I-Love-C} relations within this model are still satisfied.

%%%%%%%%%
\subsection{Gravitational binding energy relations}

\begin{figure}
    \centering
    \includegraphics[scale=0.4]{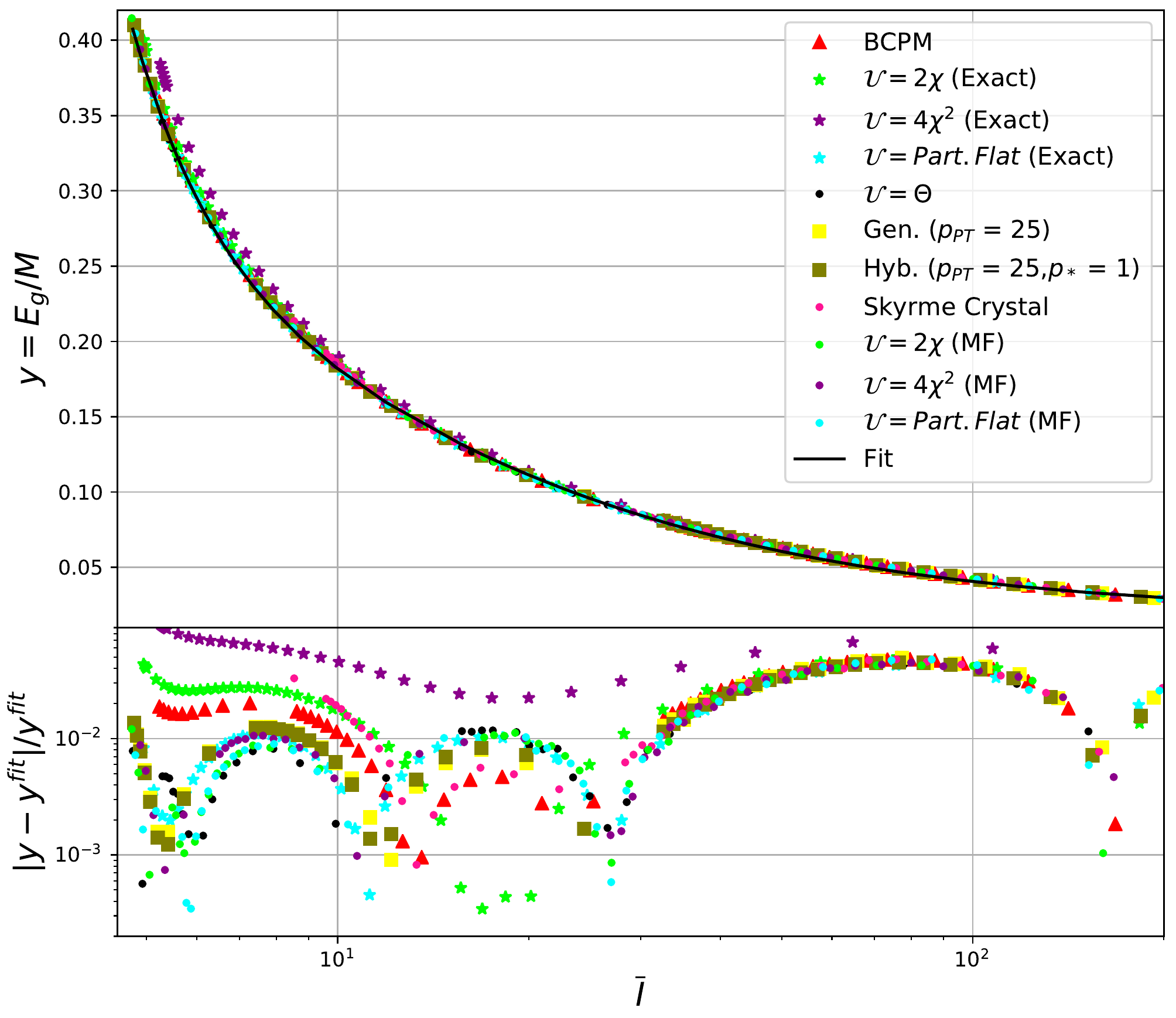}
    \caption{Relation between the adimensional moment of inertia and normalized gravitational binding energy.}
    \label{fig:EgI}
\end{figure}

\begin{figure}
    \hspace*{-0.4cm}
    \includegraphics[scale=0.4]{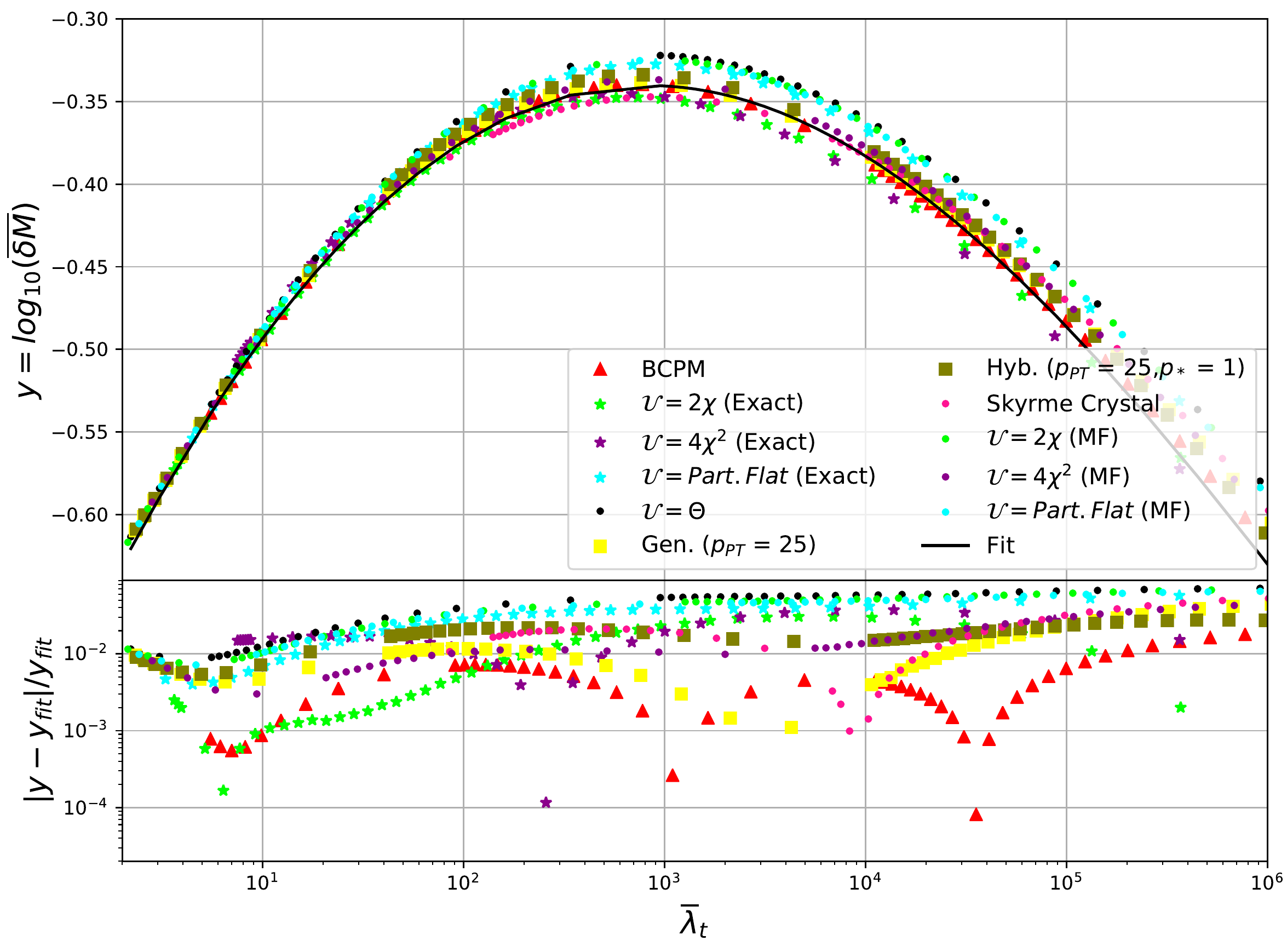}
    \caption{Quasi-universal relation for the (dimensionless) gravitational mass correction $\delta M$, and normalized deviation from the fitted relation of \cite{ReinaCompletion}.}
    \label{fig:DeltaM}
\end{figure}

A different set of quasi-universal relations involving the static gravitational binding energy and other global properties of neutron star solutions have been recently proposed in \cite{UniversalBinding}. For instance, we show in \cref{fig:EgI} the universal behavior of the static gravitational binding energy normalized to the TOV mass and plotted against the adimensional moment of inertia. From the error plot one can see that all models follow the same universal behavior with a deviation of $\lesssim 5\%$ (but the exact $4\chi^2$ BPS model, in which case the error is as high as ten percent) with respect to the numerical fit obtained in \cite{UniversalBinding}.
Further, the rotation of the star has measurable effects both in the gravitational and proper mass of the star. Indeed, as we have seen, the gravitational mass of the star receives a correction $\delta M$, 
the dimensionless version of which, $\overline{\delta M}=\delta M\times M^3/J^2$, was also shown in \cite{ReinaCompletion} to satisfy a universal relation when plotted against the (dimensionless) tidal deformability. We show this relation in \cref{fig:DeltaM}, together with the numerical fit of \cite{ReinaCompletion} obtained for the region $\overline{\lambda_t}<10^{3} $, at which the deviation for all models is less than ten percent.

\begin{figure}
    \hspace*{-0.4cm}
    \includegraphics[scale=0.4]{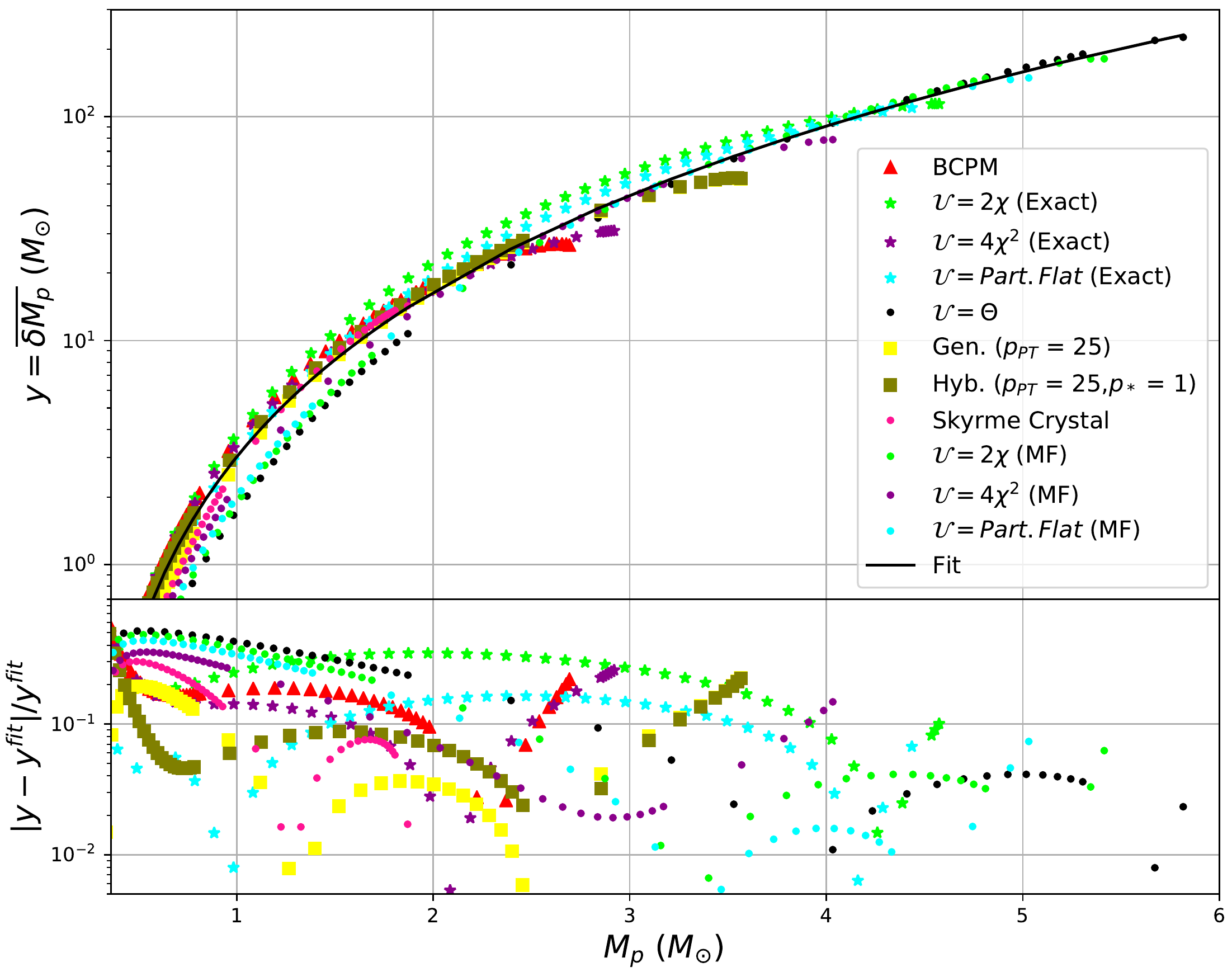}
    \caption{Normalized second order proper mass correction versus static proper mass}
    \label{fig:DeltaMp}
\end{figure}

On the other hand, the gravitational binding energy will also get a second order correction, namely, $\delta E_g$. Remarkably, as opposing to its zeroth-order counterpart, the correction to the gravitational binding energy does not seem to follow a simple, quasi-universal relation. Since the correction to the gravitational mass indeed does follow a relation as shown in \cref{fig:DeltaM}, the non-universal nature of $\delta E_g$ can be traced back to the correction to the proper mass $\delta M_p$, which involves an integral over the star, see \eqref{delMp}.  In \cref{fig:DeltaMp} we show the behavior of $\overline{\delta M_p}=\delta M_p\times M^4/J^2$ with the proper mass of the static solution, $M_p^{*}$. From this figure a curve describing a quasi-universal behavior may be inferred, which corresponds to the numerical fit we have obtained. However, this behavior has not the same universality as others previousy analyzed, as the deviation can grow up to $30\%$ for realistic masses.
\color{black}

\section{Deformability constraints from observations}
\label{sec:deform-constraints}

In addition of constituting an outstanding experimental confirmation of the validity of General Relativity, the direct observation of gravitational waves can be used to
place direct constraints on the neutron star EoS (see \cite{Nonparametric_constraints,Greif:2020pju,Constraints2020neutron} for a recent review of nuclear EoS constraints from GW observations). Indeed, the  waveform produced by the coalescence of two realistic extended bodies deviates
significantly from a point-particle waveform and thus this difference can
be observed with Advanced LIGO. The degree of the deviation, in the case of binary neutron star mergers,
depends on the underlying EoS.
Although the magnitude of the deviation is strongest at later times in
the inspiral and during the merger, Flanagan and Hinderer found that
the early phase of the inspiral depends mostly on the tidal Love number of the neutron stars, introducing a phase shift with respect to the point-particle waveform \cite{Hinderer:2009ca}.

However, the individual Love numbers for each component of the merger
cannot be separately distinguished in the observed gravitational waveform.
Instead, what can be sharply measured is the so-called effective tidal deformability, $\tilde\Lambda$ a mass-weighted average of the dimensionless deformabilities $\tilde\lambda_1$ and $\tilde \lambda_2$ of both components (with masses $m_1$ and $m_2$), given by 

\begin{equation}
    \tilde{\Lambda} \equiv \frac{16}{13} \frac{\left(m_{1}+12 m_{2}\right) m_{1}^{4} \lambda_{1}+\left(m_{2}+12 m_{1}\right) m_{2}^{4} \lambda_{2}}{\left(m_{1}+m_{2}\right)^{5}}.
    \label{effectivetidal}
\end{equation}

 Similarly, the two component masses are not measured directly. Instead, it is the chirp mass,
\begin{equation}
    M_{c}=\frac{\left(m_{1} m_{2}\right)^{3 / 5}}{\left(m_{1}+m_{2}\right)^{1 / 5}}=m_{1} \frac{q^{3 / 5}}{(1+q)^{1 / 5}},
\end{equation}
where $q=m_1/m_2$ is the mass ratio,
what can actually be constrained. In the case of the GW170817 event, the chirp mass was measured to be $1.188_{-0.002}^{+0.004}$
at the $90\%$ confidence level.  Moreover, within the same confidence level, the mass ratio was found to be in the range $0.7-1$, and the effective tidal deformability to be smaller than $800$ \cite{Abbott_2017}.

Such such measurements of the NS properties can be used to further reduce the set of Skyrme models able to reproduce physically realistic NS solutions and impose some constraints on the possible values of the free parameters of these models. Indeed, once the equations for the tidally deformed stars of \cref{Sec:Tid.demform} are solved for a specific model, we may obtain the dimensionless tidal deformability of stars described by this model as a function of their TOV mass, so that $\tilde \Lambda$ may be seen as a function of both $m_1$ and $m_2$, or, equivalently, of $M_c$ and $q$. On the other hand, since the chirp mass
of the binary progenitor of GW170817 is well measured,
for any given EOS the effective deformability reduces to a simple EOS-dependent function of the mass ratio.

\begin{figure}[h!]
%\hspace*{-0.5cm}
    \includegraphics[scale=0.37]{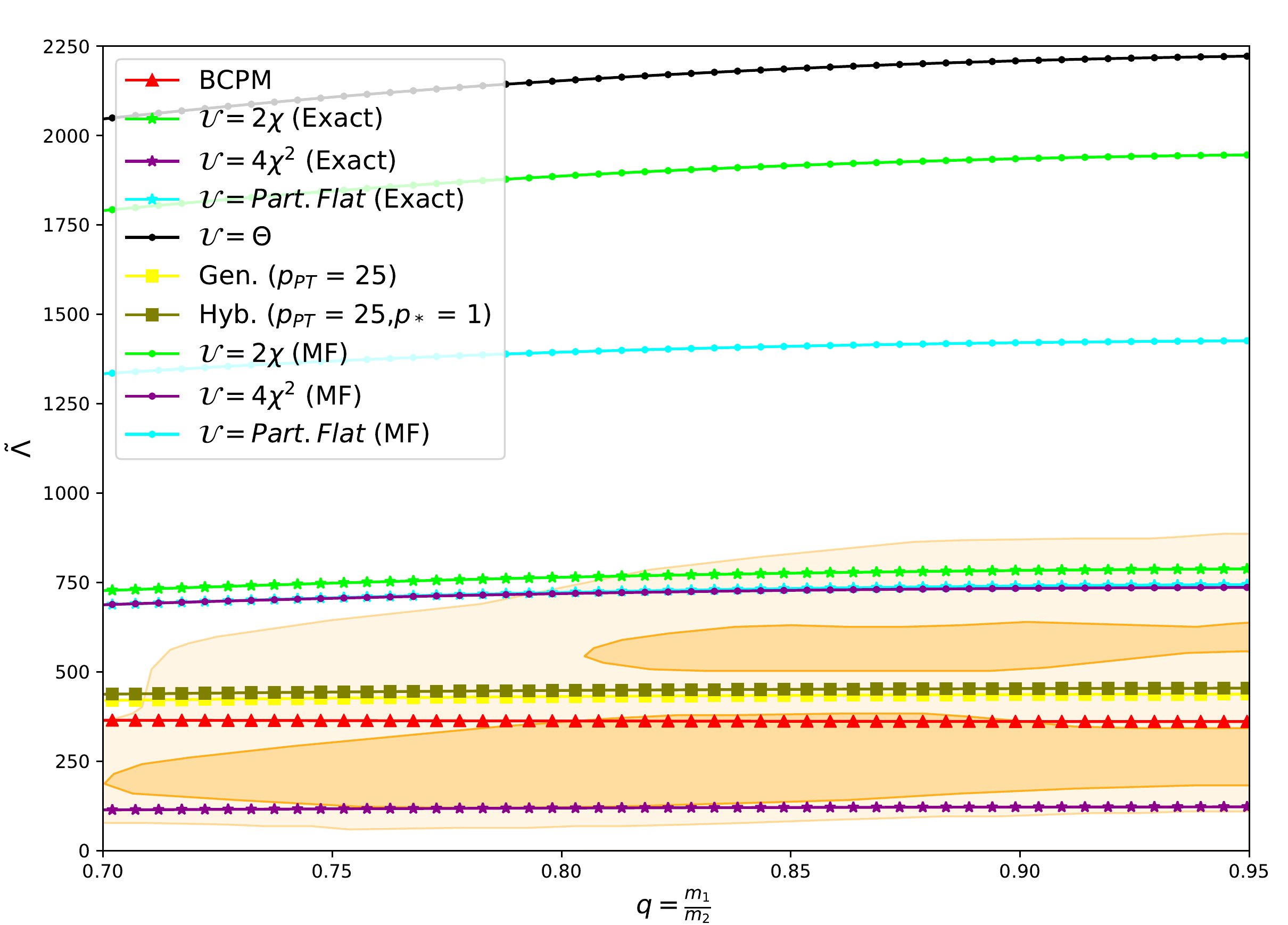}
    \caption{Effective tidal deformability versus mass ratio of the two merging stars.}
    \label{fig:LAMBDATIDAL}
\end{figure}

In \cref{fig:LAMBDATIDAL} we show the effective tidal deformability as a function of the mass ratio for a chirp mass of 1.19 $M_\odot$ for different Skyrme models, together with the constraints from the GW170817 event. It is clear from this figure that, as we have already argued,  the mean field approximation is not suitable for describing the low energy region of BPS stars, which makes the most relevant contribution to the deformability of the stars. Indeed, this approximation overestimates the values of effective deformability by at least a factor of $\sim 2$.

 On the other hand, we find that both the generalized and the hybrid EoS provide an excellent description of the tidal deformability \cite{Adam:2020yfv}.

In addition, we see again that the exact BPS Skyrme models present very different behaviors ---in this case, different values of $\tilde\Lambda$--- depending on the chosen potential. For example, the contribution from the $\Theta$ potential is clearly too big as compared with the GW observation, which sets an upper value of $\Tilde{\Lambda}\leq 800$, whereas the $2\chi$ and PF potentials yield very large $\tilde\lambda$, near the upper value, although still allowed by the bound).

Furthermore, it is likely that additional observations of gravitational waves from binary NS mergers will further constrain the tidal deformability of these compact stars. In particular, some recently observed GW events \cite{Abbott:2020uma,Abbott:2020khf} strongly suggest that highly massive NS and compact objects within the NS-Black Hole mass gap (around $2.5\,M_\odot$)  could exist. 

However, it is difficult to distinguish between an extremely massive neutron star and a small black hole from the GW waveform alone with first generation GW detectors, since the tidal deformability and quadrupole moment of such massive stars is usually very low due to their high compactness, and almost no realistic EoS is able to produce stars with such big mass.
\begin{figure}
    \hspace*{-0.7cm}
    \includegraphics[scale=0.35]{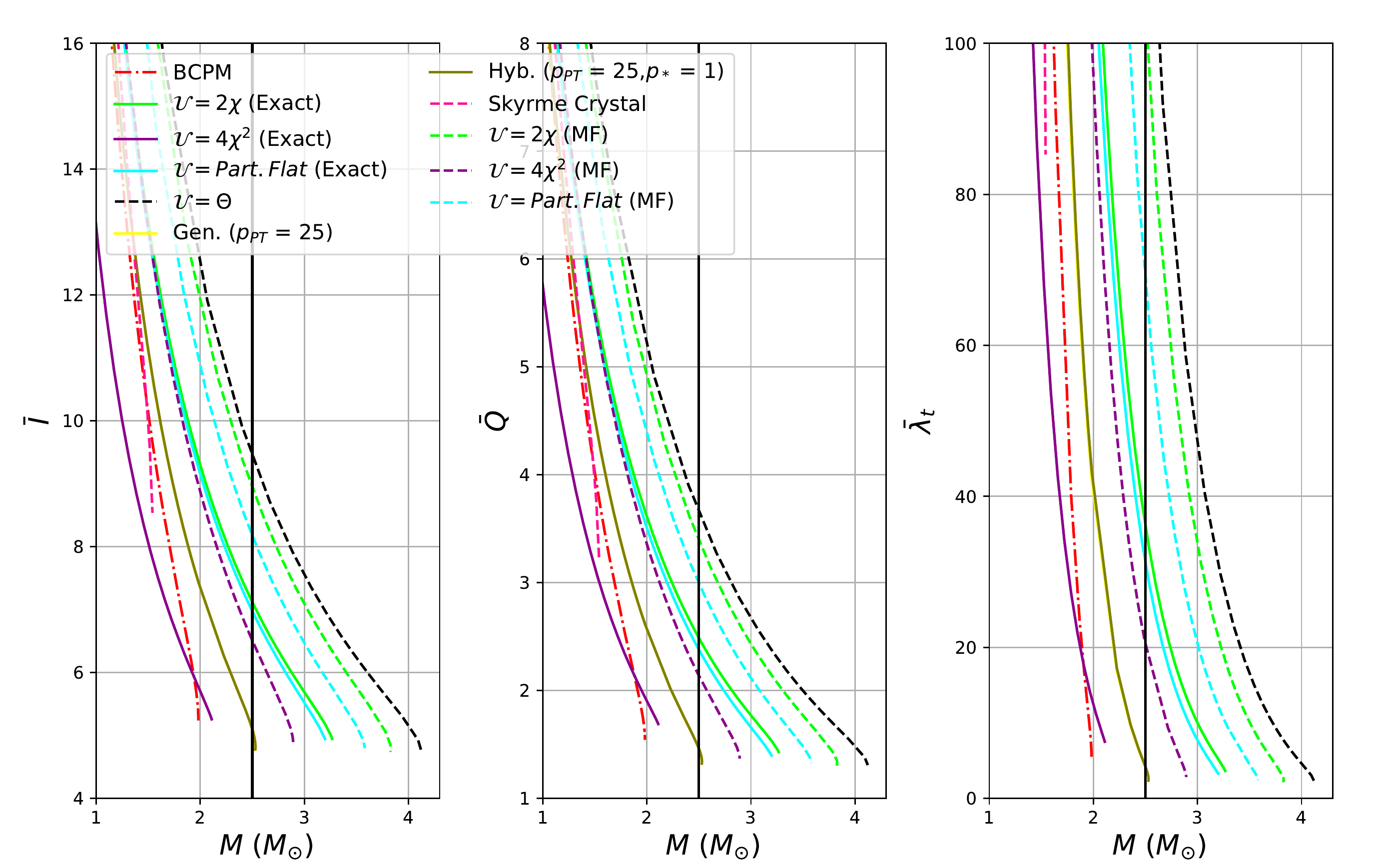}
    \caption{Dimensionless moment of inertia, quadrupolar moment and tidal deformability versus mass of stars for different models and EoS}
    \label{fig:IQL}
\end{figure}

In \cref{fig:IQL}, we show the dimensionless moment of inertia, quadrupolar moment and tidal deformability of all the Skyrme models as well as for the AP4 and BCPM EoS. These plots show that not only high mass neutron star solutions can be found for any BPS Skyrme model as well as for the generalized and the hybrid EoS. We also find that, depending on the potential, these parameters can acquire sufficiently high values to be able to be measured by current generation GW observatories. Therefore, we conclude that if the tidal deformability of a mass-gap compact object were measured to be non-zero, it is very likely that its EoS will be well approximated by a BPS Skyrme model and, in particular, by the hybrid model which approaches the BPS behavior at high density.

\section{Conclusions}

In this work, we have solved the Einstein equations using the Hartle-Thorne perturbative formalism to find slowly rotating NS solutions with nuclear and Skyrme model-based EoS. Moreover, we have presented perturbative solutions to the Einstein-BPS Skyrme system describing slowly rotating and tidally deformed, self-gravitating solitons which can also be considered as idealized models for Neutron Stars. For all these models, we have computed different global properties of the corresponding star configurations, such as the moments of inertia, quadrupole moments, gravitational masses or binding energies, and checked whether or not all the models satisfy some (quasi-)universal relations previously proposed in the literature. As we have found, the \emph{I-Love-Q} relations presented in \cite{Yagi_2013} are satisfied up to a $\sim 2\%$ error, even for the exact, non-barotropic BPS Skyrme models, which reaffirms the universality of these relations. Other relations involving the second-order correction to the gravitational mass (including the correction proposed in \cite{ReinaCompletion}) and those involving the (gravitational) binding energy are also quite well satisfied for all models at hand.

On the other hand, we have found that while the \emph{I-Love-C} quasi-universal relations still hold for the mean-field BPS and Skyrme-based EoS, these relations break up for the exact BPS Skyrme models. This fact, as argued, can be traced back to the behavior of the energy density profiles of the solutions for such models, which strongly depends on the particular potential chosen due to the non-barotropic nature of these models. This finding is consistent with the explanation given in \cite{Yagi:2014qua} about the difference in nature between these relations and the \emph{I-Love-Q}.

The extension of previous works on Skyrmion stars to include the effects of small rotations and tidal deformations allows to enlarge the set of observable quantities that can be compared to actual measurements. 
Owing to the increasing number of observed GW events in recent years, those observables that can be inferred from the waveform of a GW produced at a binary NS coalescence are of particular interest.
An example is the (effective) tidal deformability of the binary, which, together with the quasi-universal relations, allows to constrain the EoS of strongly interacting matter in the extremely high density regime. In this paper, we have shown that, within the Skyrme model, these universal relations still hold and that the current experimental bounds on NS deformabilities can be well accounted for. 
Furthermore, a remarkable property of the solutions based on generalized Skyrme models is that very high masses (of approximately $2.5 M_\odot$) can be reached even for not too large energy densities at the center of the stars. In other words, such massive stars can be produced from mesonic degrees of freedom alone without the need of additional degrees of freedom such as unconfined quarks. This is consistent with the assumption that the Skyrme model is a valid approximation for the description of matter at the core of a NS, which, if true, implies that the pressure and density reached at NS cores are still far from the energy density regimes in which perturbative QCD becomes relevant.

Although some recent GW events can be seen as possible evidence that such massive stars may exist, additional observations are required to further clarify the detailed properties of massive NS cores.

We conclude by summarizing our main results:

{\em i)} we find that the hybrid EoS, where the EoS of the generalized
Skyrme model is complemented by a standard nuclear physics EoS for low
densities, is compatible with {\em all} observational constraints, both
for static NS observables and for observables related to slowly rotating and/or tidally deformed
NS, and

{\em ii)} we verify (quasi-)universal relations, like {\em I-Love-Q}, for
a broad range of models, based on the minimal Skyrme model, the BPS Skyrme
model with a variety of potentials, the generalized Skyrme model, and the
hybrid model, respectively.
In particular, the BPS Skyrme model also allows for an exact
field-theoretic treatment (beyond mean-field theory), because it
represents a non-barotropic perfect fluid.
These results contribute to a deeper understanding of the range of validity of these
relations, because we investigate them for qualitatively different models
not considered previously.
In addition, our investigation also provides a better insight into the
role played by each component of the full, hybrid Skyrme model.

\begin{acknowledgements}
The authors acknowledge financial support from the Ministry of Education, Culture, and Sports, Spain (Grant No. FPA2017-83814-P), the Xunta de Galicia (Grant No. INCITE09.296.035PR and Conselleria de Educacion), the Spanish Consolider-Ingenio 2010 Programme CPAN (CSD2007-00042), Maria de Maetzu Unit of Excellence MDM-2016-0692, and FEDER.
AGMC is also grateful to the Spanish Ministry of Science, Innovation and Universities, and the European Social Fund for the funding of his predoctoral research activity (Ayuda para contratos predoctorales para la formación de doctores 2019).
\end{acknowledgements}

\appendix
\section{Linear perturbations of spherically symmetric spacetimes}
\label{Appendperturb}
In this appendix we establish the notation for the gravitational perturbations of Schwarzschild spacetime as introduced in \cite{Regge:1957td}, and developed in \cite{ Martel_2005}.

Consider a spherically symmetric static spacetime $\mathcal{M}$, whose metric is given in general by the Schwarzschild solution, whose line element, can be written:
\begin{equation}
    ds^2=g_{ab}dx^adx^b+r^2(x^a)\Omega_{AB}dx^Adx^B
\end{equation}
with lower-case latin indices running over $\qty{0,1}$ and upper-case latin indices running over $\qty{2,3}$. In particular, $r$ is a scalar function of the lower-case coordinates, the coordinates $x^A=\qty{\theta,\phi}$ span the two-spheres ${x^a=\text{const}}$, and $\Omega_{AB}$ is the metric on the unit two-sphere, ${\Omega_{AB}=\text{diag}(1,\sin^2\theta)}$. Let $D_A$ be the covariant derivative operator compatible with $\Omega_{AB}$, and $\varepsilon_{AB}$ the Levi-Civita tensor on the unit two-sphere, with $\varepsilon_{\theta\phi}=\sin\theta$. 

We may now introduce the metric perturbation $\delta g_{\mu\nu}:=h_{\mu\nu}$, which may be written
\begin{equation}
    h_{\mu\nu}=h_{ab}dx^adx^b+h_{aB}dx^adx^B+h_{AB}dx^Adx^B.
    \label{metricpertcov}
\end{equation}
Spherical symmetry of the background spacetime motivates a decomposition of $h_{\mu\nu}$ in spherical harmonics. Note that the $h_{ab}$ transform as scalars on $\mathcal{S}^2$, whereas $h_{aB}$
 and $h_{AB}$ transform as covariant vectors and tensors, respectively, on $\mathcal{S}^2$. Therefore, we should decompose them into scalar, vector and tensor spherical harmonics, respectively.
 
 The scalar harmonics are the usual spherical-harmonic functions $Y^{lm}(x^A)$, which satisfy the eigenvalue equation:
 \begin{equation}
     [\Omega^{AB}D_AD_B+l(l+1)]Y^{lm}=0.
 \end{equation}
Vectorial spherical harmonics come into two types, depending on their parity:
\begin{enumerate}[a)]
    \item Even parity: $Y^{lm}_A:=D_A Y^{lm}$
    \item Odd parity: $X^{lm}_A := -\varepsilon_A^{\,B}D_BY^{lm}$
\end{enumerate}
Tensorial harmonics are also classified according to their parity:
\begin{enumerate}[a)]
    \item Even parity: there are two kinds: \[\left\{
    \begin{array}{ccll}
         Y^{lm}_{AB}&:=&\Omega_{AB}Y^{lm},\,\text{(traceful, scalar degree of freedom)}\\\\
         \Tilde{Y}^{lm}_{AB}&:=&[D_AD_B-\tfrac{1}{2}l(l+1)\Omega_{AB}]Y^{lm}\,\text{(traceless)} 
    \end{array}
    \right.\]
    \item Odd parity: $X^{lm}_{AB}:=\frac{1}{2}(\varepsilon_A^{\,C}D_B+\varepsilon_B^{\,C}D_A)D_CY^{lm}$.
\end{enumerate}
Therefore, the components of the metric perturbation \eqref{metricpertcov} can be written:
\begin{equation}
    \begin{split}
        h_{ab}=&\sum\limits_{l,m}h_{ab}^{lm}(x^a)Y^{lm}(x^A)\\
        h_{aA}=&\sum\limits_{l,m}j_{a}^{lm}(x^a)Y_A^{lm}(x^A)+\sum\limits_{l,m}h_{a}^{lm}(x^a)X_A^{lm}(x^A)\\
        h_{AB}=&\sum\limits_{l,m}r^2(x^a)\left[k^{lm}(x^a)Y_{AB}^{lm}(x^A)\right.\\&\quad+G^{lm}(x^a)\tilde Y^{lm}_{AB}(x^A)+h_2^{lm}(x^a)X^{lm}_{AB}(x^A)]
    \end{split}
    \label{Metricperturbationgeneral}
\end{equation}

The even-parity sector of the perturbation (also called polar perturbations) consists of the associated functions $h_{ab}^{lm},j_a^{lm},k^{lm}$ and $G^{lm}$, whilst the variables $h_a^{lm}$ and $h_2^{lm}$ make up the odd parity sector (axial perturbations).

As noted in the main text, a well defined perturbative treatment of spacetime introduces a gauge freedom.
Up to now, the discussion of the perturbation functions have been made in an arbitrary gauge. It is useful to fix the gauge for the metric perturbations, in order to simplify the problem of determining their explicit expression by solving the perturbed Einstein equations. 
A useful gauge choice is the so-called Regge-Wheeler gauge \cite{Martel_2005}, in which
$
   { j^{lm}_a=G^{lm}=h_2^{lm}=0.}
$

Furthermore, for stationary, axially symmetric perturbations, we may discard the $\phi-$dependence of the harmonics, and the only nonvanishing contribution will be that of the $m=0$ terms. Hence, the general expression for an stationary, axially symmetric metric perturbation with these gauge choices is given by $h_{\mu\nu}= h^\text{even}_{\mu\nu}+ h^\text{odd}_{\mu\nu}$, where:

\begin{align}
    h^\text{even}_{\mu\nu}=&\mqty(-H_l&I_l&0&0\\
                    I_l&M_l&0&0\\
                    0&0&r^2K_l&0\\
                    0&0&0&r^2\sin^2(\theta)K_l)Y^{l0}(\theta)\\
                    h^\text{odd}_{\mu\nu}=&\mqty(0&0&0&\omega_l\\
                                0&0&0&V_l\\
                                0&0&0&0\\
                            \omega_l&V_l&0&0)X_\phi^{l0}(\theta)
    \label{MetricperturbationPleg}
\end{align}
where $Y^{l0}=P_l(\cos\theta)$ and ${X_\phi^{l0}=\sin{\theta}\partial_\theta P_l(\cos\theta)}$, and summation over $l$ is implied.

\vspace*{0.5cm}

%\nocite{*}
\bibliography{biblio}% Produces the bibliography via BibTeX.
\end{document}